\documentclass[epj,usenames,dvipsnames]{svjour}
\pdfoutput=1
\usepackage{amsmath}
\usepackage{colortbl}
\usepackage{xcolor}
\usepackage{hyperref}
\usepackage{nccmath}
\usepackage{cite}

\newcommand{\beq}{\begin{equation}}
\newcommand{\eeq}{\end{equation}}

\newcommand{\neut}{$\textsc{neut}$}
\newcommand{\genie}{$\textsc{genie}$}

\usepackage{array, graphicx, subfigure}
\usepackage{graphics}
\begin{document}
%
\title{Comparison of optical potential for  nucleons and  $\Delta$ resonances
}
\subtitle{In electron scattering from nuclear targets }
\author{Arie Bodek 
 and Tejin Cai
 }
\institute{Department of Physics and Astronomy, University of
Rochester, Rochester, NY  14627-0171}
\date{ arXiv:2004.00087 Version 5  July 4,  2020 to be published in EPJC}  
%
\abstract{
  Precise modeling of  neutrino  interactions on nuclear targets is essential for neutrino oscillations experiments. The modeling of  the energy of final state particles in quasielastic (QE) scattering  and resonance production  on bound nucleons requires knowledge of both the removal energy of the initial state bound  nucleon as well  as the average Coulomb and nuclear optical potentials for final state leptons and hadrons. We extract the average values of the real part of the nuclear optical potential for final state nucleons ($U_{opt}^{QE}$)  as a function of the nucleon kinetic energy from  inclusive electron scattering data on nuclear targets ($\bf_{6}^{12}C$+$\bf_{8}^{16}O$, $\bf_{20}^{40}Ca$+$\bf_{18}^{40}Ar$, $\bf_{3}^{6}Li$,  $\bf_{18}^{27}Al$, $\bf_{26}^{56}Fe$, $\bf_{82}^{208}Pb$)   in the QE region and compare to calculations. We also extract values of the  average of the real part of the nuclear optical potential for a $\Delta(1232)$ resonance in the final state ($U^\Delta_{opt}$) within the impulse approximation.   We find that  $U^\Delta_{opt}$ is more negative than $U_{opt}^{QE}$  with $U^\Delta_{opt}\approx$1.5~$U_{opt}^{QE}$ for  $\bf_{6}^{12}C$.
} 
%
\maketitle
%

\section{Introduction}
\label{intro}
Precise modeling of  neutrino  interactions on nuclear targets is essential for neutrino oscillations experiments\cite{MINOS,NOVA,K2K,MiniBooNE,DUNE}.
The modeling of  the energy of final state particles in quasielastic (QE) scattering  and resonance production  on bound nucleons requires knowledge of both the removal energy of the initial state bound  nucleon as well  as the average Coulomb and nuclear optical potentials for final state leptons and hadrons. In this communication we compare the values of the average nuclear optical potential for final state nucleons ($U_{opt}^{QE}$) as a function of the nucleon kinetic energy extracted from inclusive electron scattering data on nuclear targets 
in the QE region to calculations based on proton scattering data (on nuclear targets).   In addition, we compare to values of the average nuclear optical potential for a $\Delta(1232)$ resonance in the final state ($U^\Delta_{opt}$) extracted from a subset of the inclusive electron scattering data. 

First we summarize some of the results of our  previous publication\cite{optpaper} on removal energies and the average
  nuclear optical potential for final state nucleons  extracted from inclusive quasielastic (QE) electron scattering data  on a variety of nuclei.
  The analysis was done within the framework of the impulse approximation.  
  
\begin{figure*}
\begin{center}
\includegraphics[width=3.5in,height=2.6in]{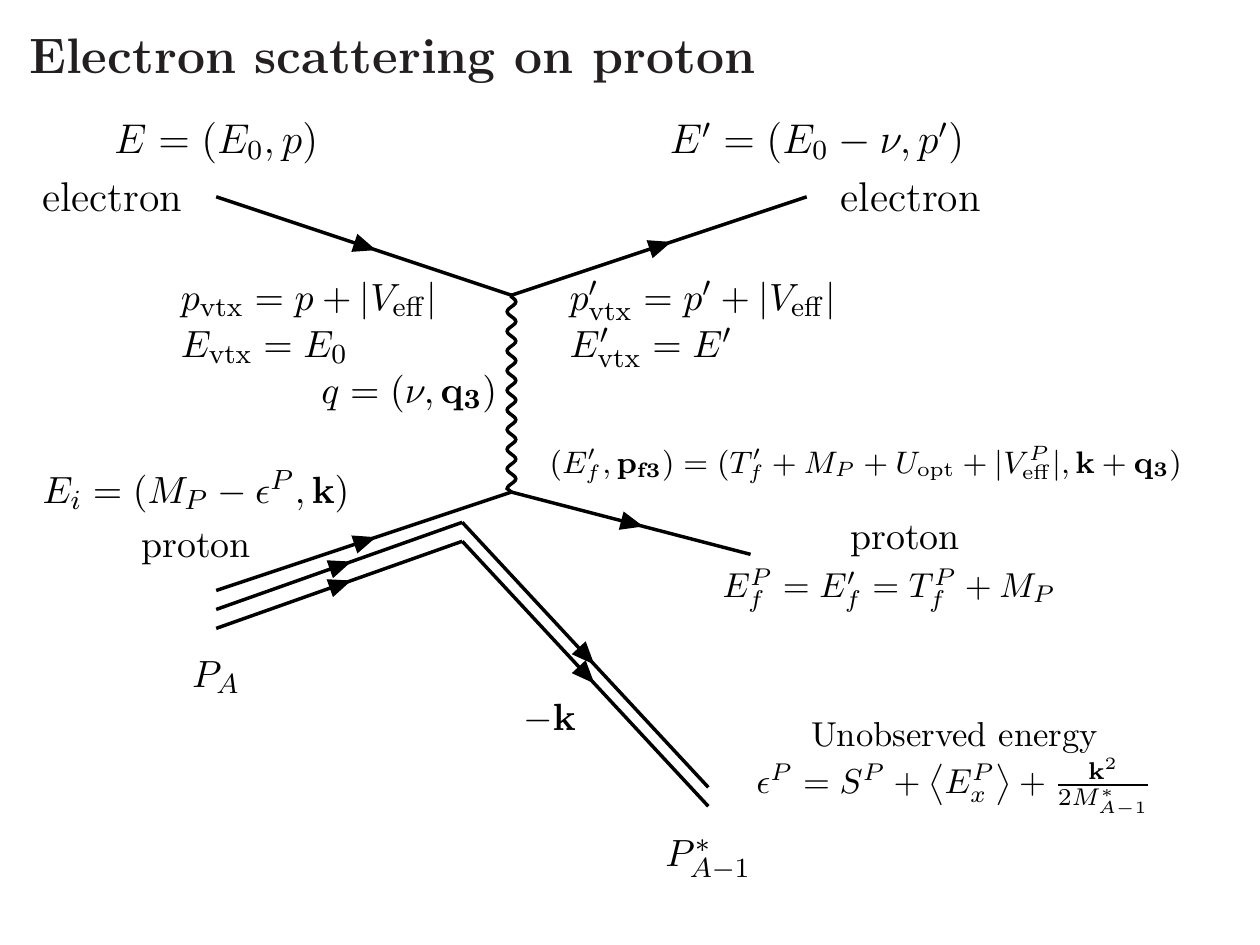}
\includegraphics[width=3.5in,height=2.6in]{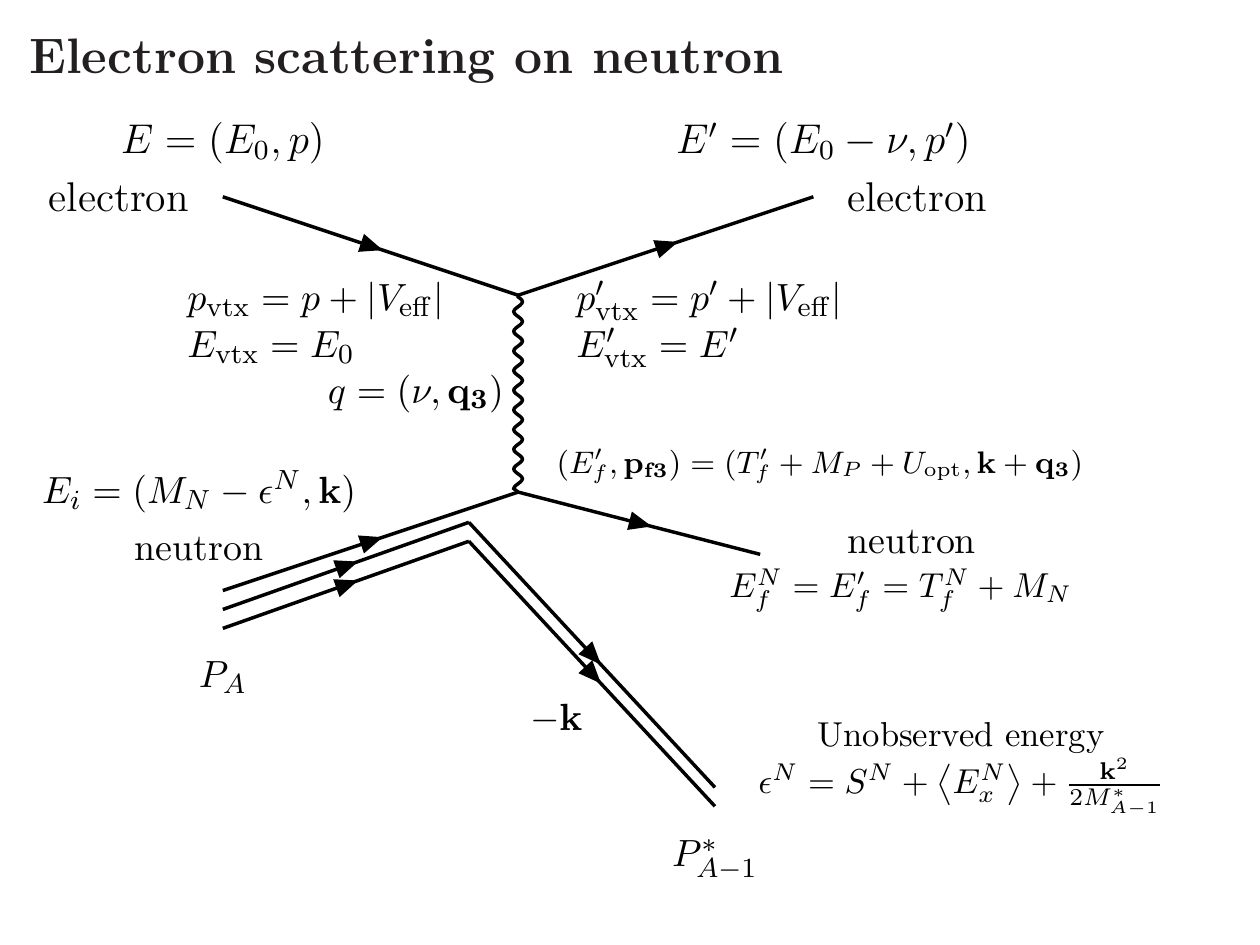}
\includegraphics[width=3.5in,height=2.6in]{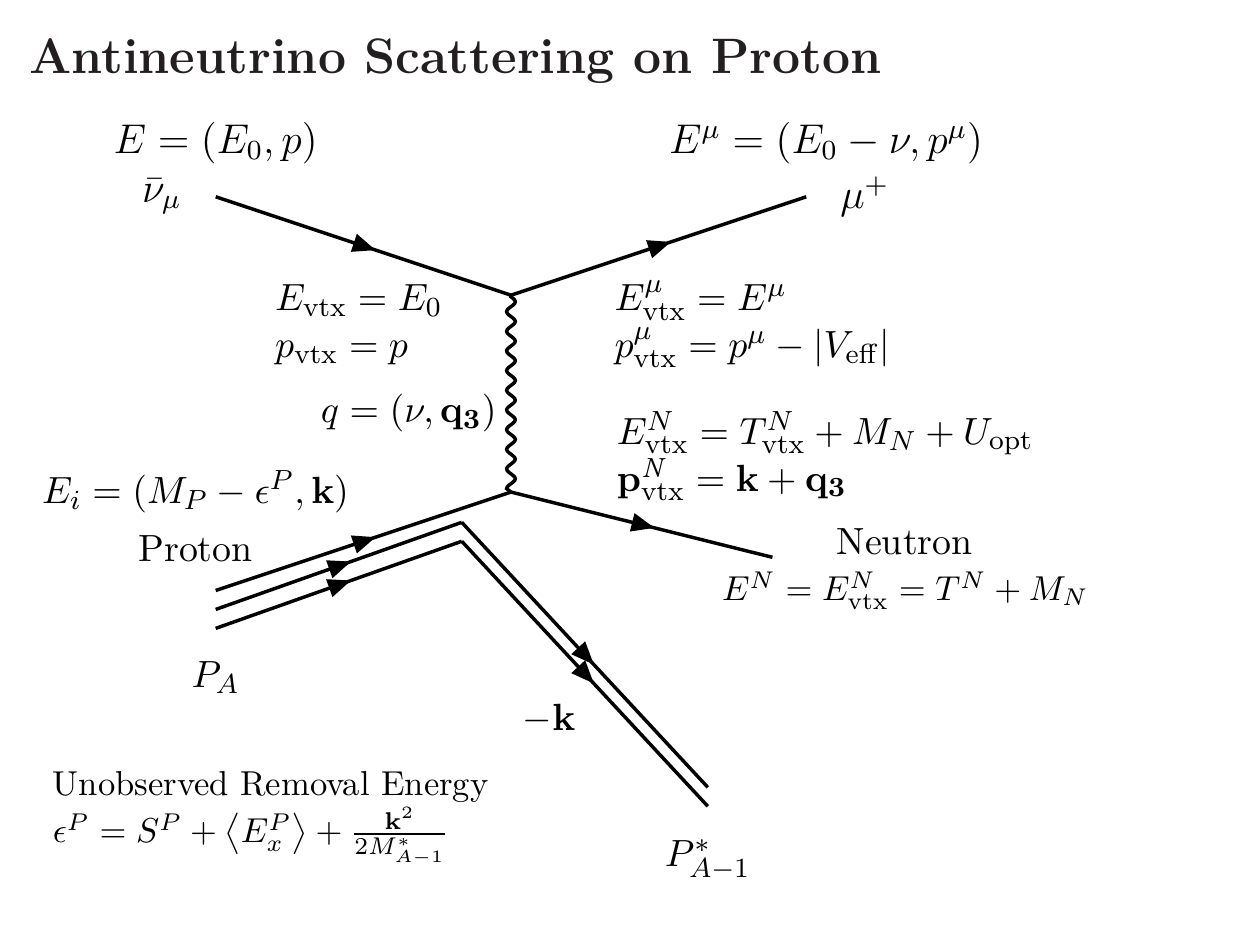}
\includegraphics[width=3.5in,height=2.6in]{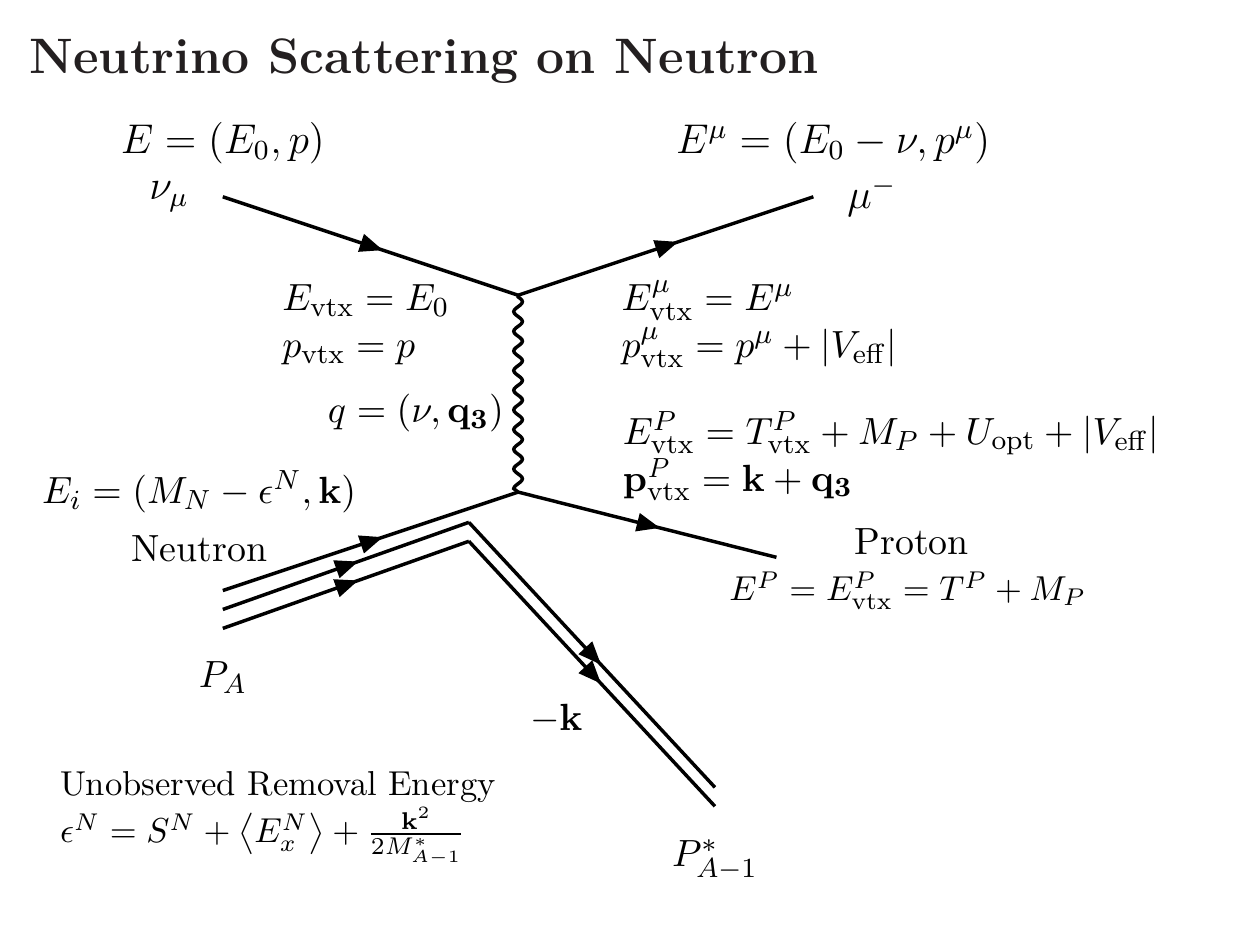}
\caption{ The diagrams on the top two panels show electron QE  scattering from an off-shell bound proton (left) and neutron (right). The diagrams on the bottom two panels show $\bar\nu$ QE scattering from an off-shell bound proton producing a final state neutron (left), and $\nu$ scattering from an off shell bound neutron producing a final state proton (right).}
\label{QEdiagram}
\end{center}
\end{figure*}

  The diagrams on the top two panels of Fig. \ref{QEdiagram} illustrate 
   electron QE  scattering from an off-shell bound proton (left) and neutron (right). The diagrams on the bottom two panels show
antineutrino ($\bar\nu$) QE scattering from an off-shell bound proton producing a final state neutron (left), and neutrino ($\nu$) scattering from an off shell bound neutron producing a final state proton (right).
The electron scatters from an off-shell nucleon of momentum $\vec {p_i}$=$\vec k$ bound in  a nucleus of mass A.  For electrons of incident energy $E_0$ and final state energy $E^\prime$,  the energy transfer to the target is $\nu=E_0-E^\prime$.  The square of the 4-momentum transfer ($Q^2$),  and 3-momentum transfer ($\vec q_3$) to a nucleon bound in the nucleus are: 
  \begin{eqnarray}
       Q^2&=& 4(E_0+|V_{eff}|)(E_0-\nu+|V_{eff}|)\sin^2\frac{\theta}{2}\\
         {\vec {q_3^2}}&=&Q^2+\nu^2. \nonumber
   \end{eqnarray}
   We include the effects  of the interaction of  initial and final state electrons with the Coulomb field of the nucleus by using published values of the average Coulomb energy at the interaction vertex $V_{eff}$ extracted from a comparison of  electron and positron inclusive QE differential  cross sections\cite{veff}.  These values are in agreement with calculations based on charge distributions of nuclei\cite{charge}. In the diagrams of  Fig. \ref{QEdiagram}, the energies shown include both kinetic and potential energies.
   \begin{figure*}
  \centering
 \includegraphics[width=0.42\textwidth]{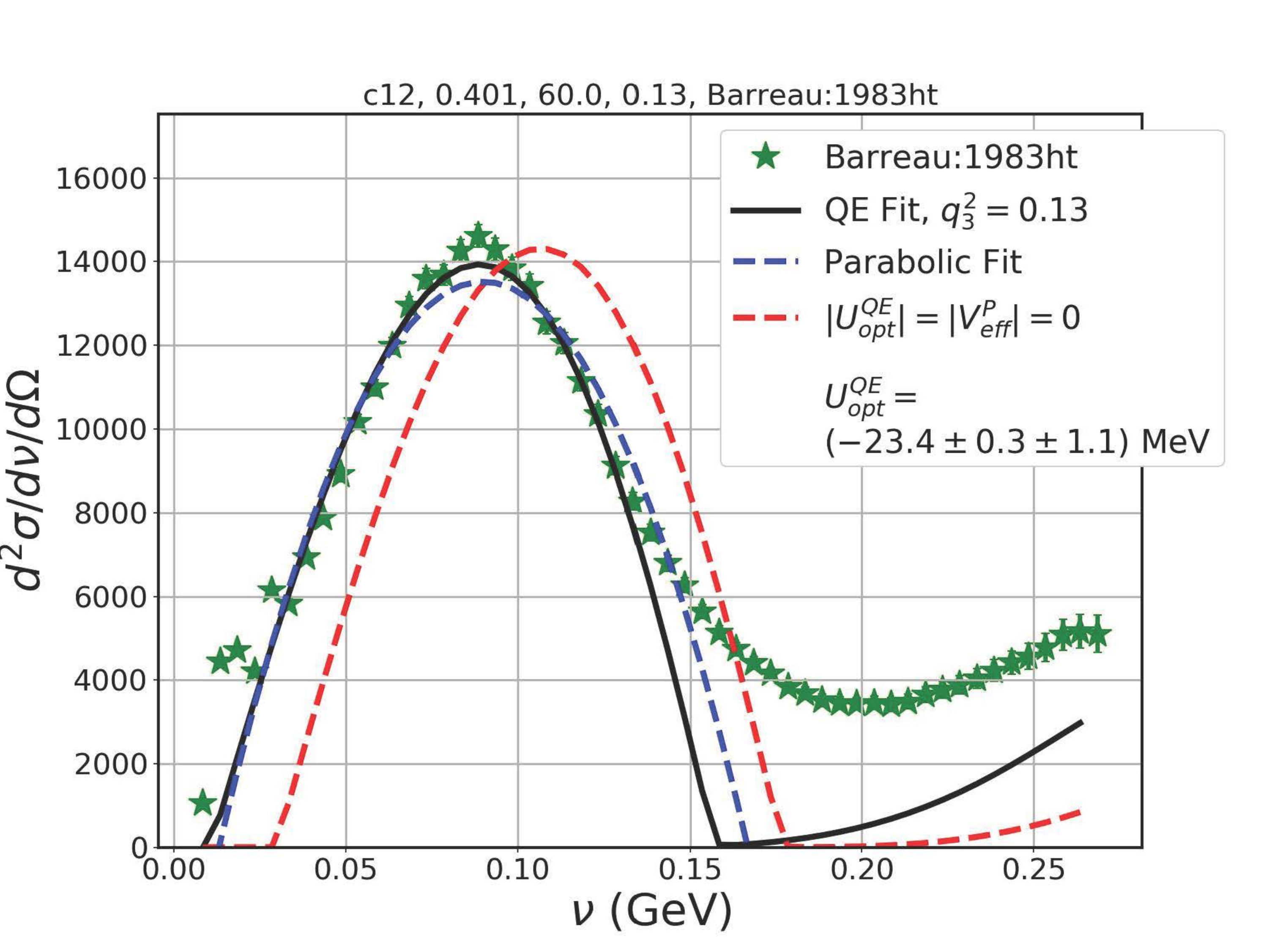}
 \includegraphics[width=0.42\textwidth]{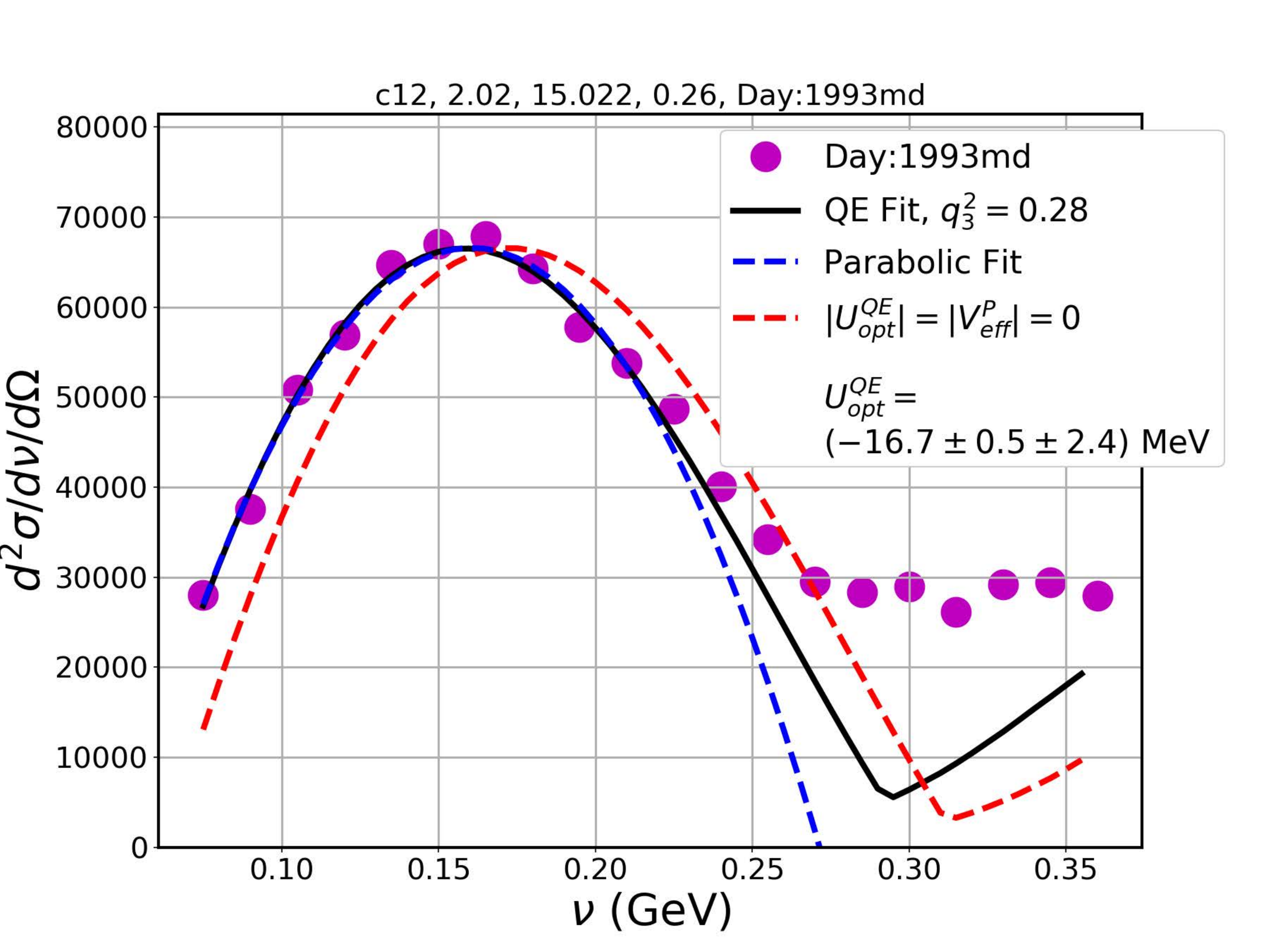}
\caption{
\footnotesize\addtolength{\baselineskip}{-1\baselineskip} 
Examples of fits for  two out of  33  $\bf_{6}^{12}C$ QE differential cross sections. The solid black curves are the RFG fits with the best value of $U^{QE}_{opt}$ for the final state nucleon. The blue dashed curves are simple parabolic fits used to estimate the systematic error.  The difference between $\nu_{peak}^{parabola}$ and $\nu_{peak}^{rfg}$ is used  as a systematic error in our  extraction of $U^{QE}_{opt}$.   The first error shown in the legend is the statistical error in the fit.  The second error is the systematic uncertainty which is much larger. The red dashed curve is the RFG model  with  $U^{QE}_{opt}=0$ and $|V^P_{eff}|=0$.
}
\label{C12_fits}
\end{figure*}
\begin{figure*} 
\centering
      \includegraphics[width=0.7\textwidth]{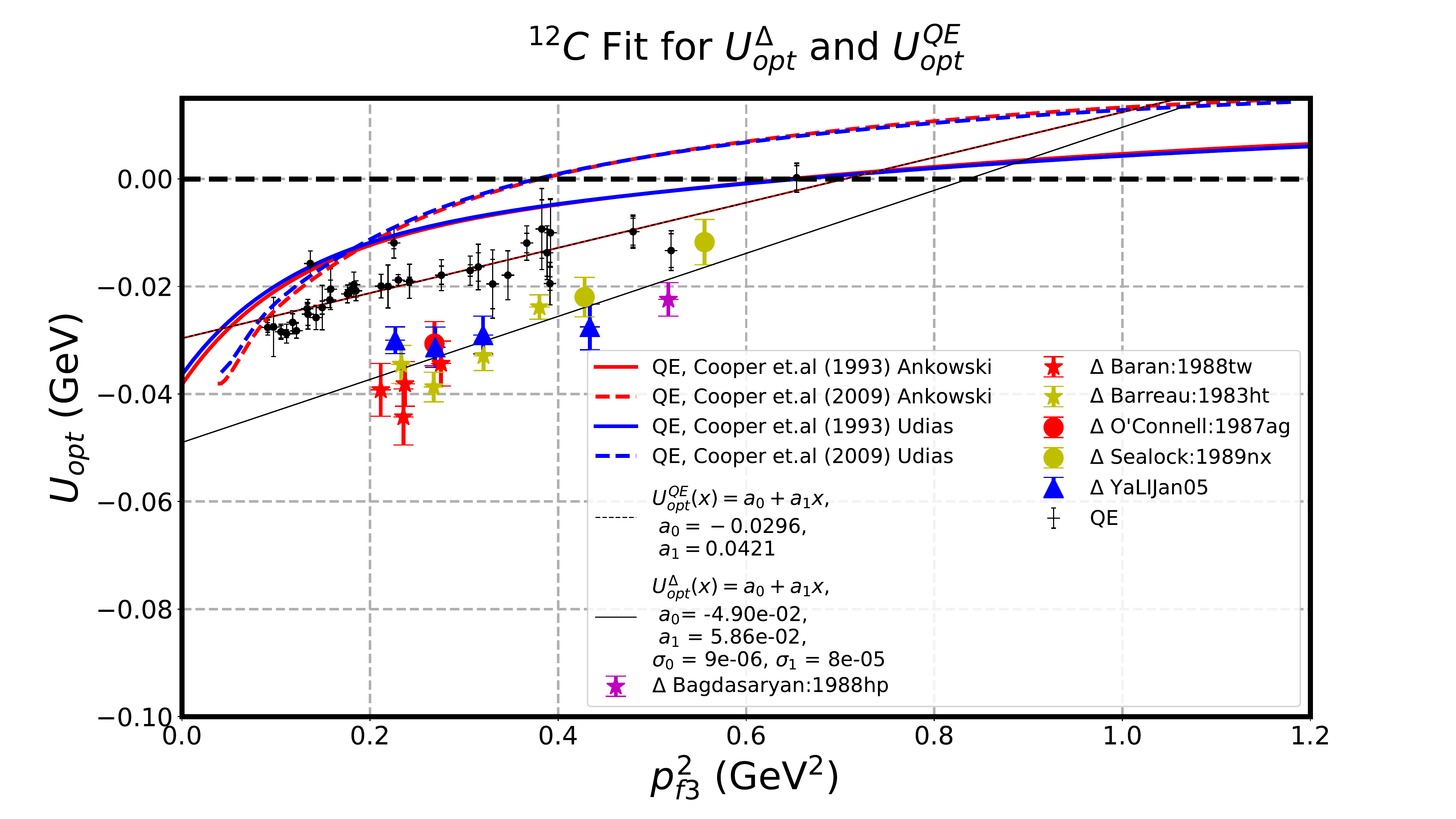}
    \includegraphics[width=0.7\textwidth]{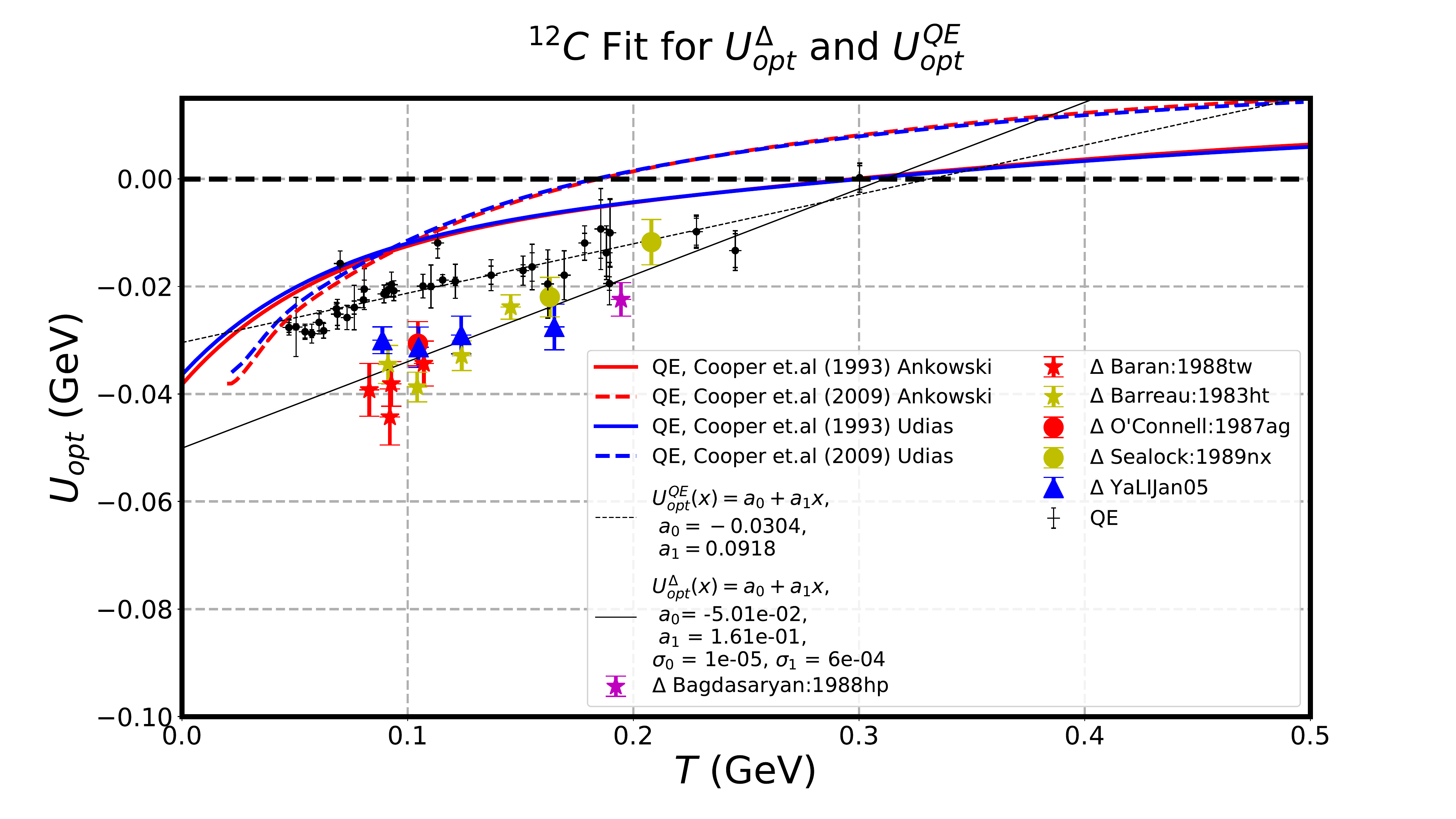}
  \caption{
\footnotesize\addtolength{\baselineskip}{-1\baselineskip} 
 Extracted values of $U^{QE}_{opt}$  for the final state nucleon in QE scattering (small black markers) for  33 $\bf_{6}^{12}C$ and four $\bf_{8}^{16}O$ inclusive electron scattering  spectra.  Also shown are  prediction for $U^{QE}_{opt}$  calculated by and Jose Manuel Udias\cite{Udias} and Artur. M. Ankowski\cite{Artur}  using the theoretical formalisms of  Cooper 1993\cite{Cooper1993} and Cooper 2009\cite{Cooper2009}.   The dashed grey lines are linear fits to the QE data.
  The larger markers are the  values of  $U^{\Delta}_{opt}$ for the final state $\Delta$(1232) extracted from a subset of the data  (15  $\bf_{6}^{12}C$ spectra)  for which the measurements extend to higher invariant mass. Here, the solid grey lines are linear fits to the $U^{\Delta}_{opt}$ values. The top and bottom panels show the measurements  versus $\vec p_{f3}^2= (\vec k +\vec{q}_3)^2$, and versus hadron kinetic energy T, respectively.  }
  \label{UC12}
\end{figure*}
%
\begin{figure*} 
\centering
             \includegraphics[width=0.65\textwidth]{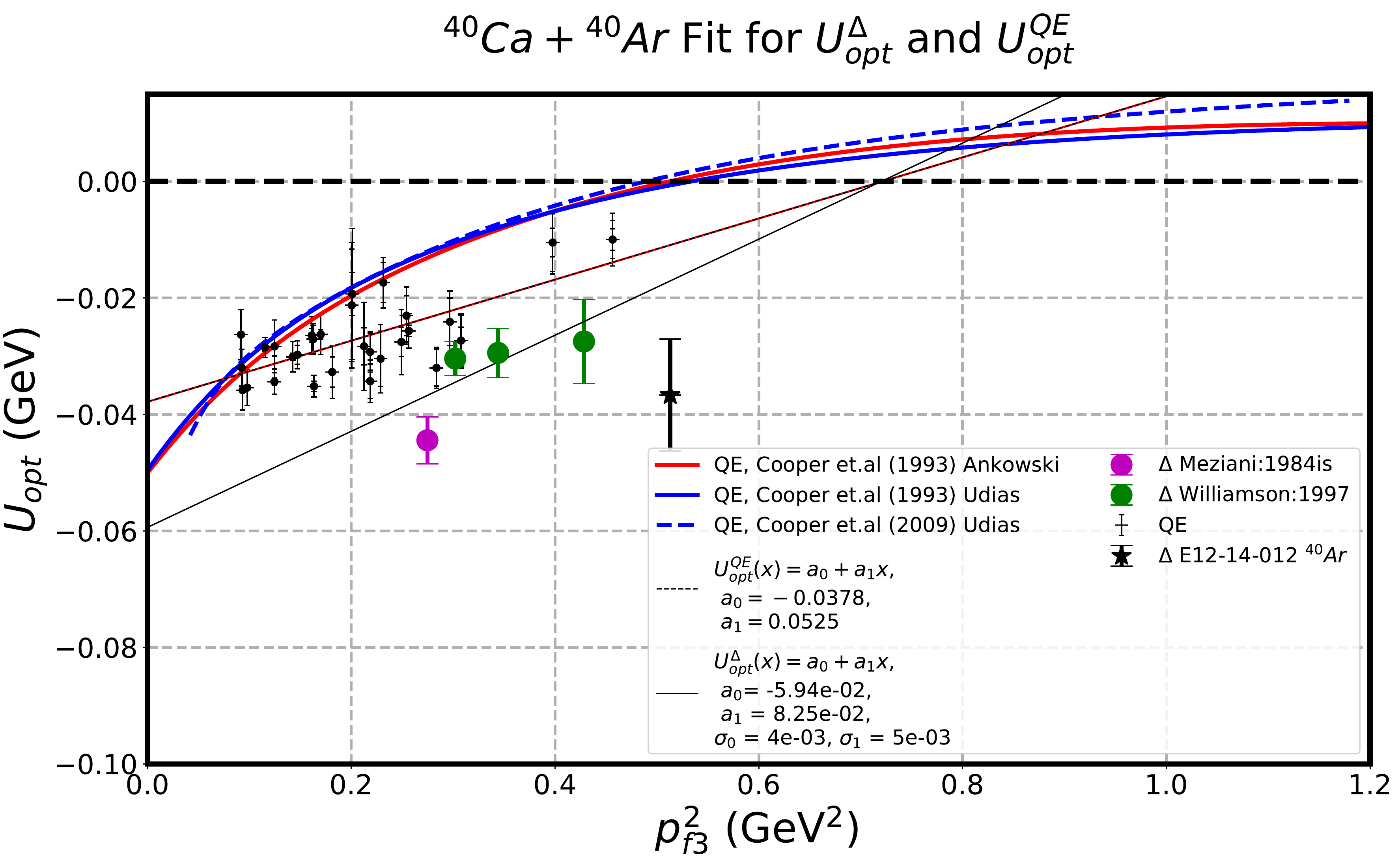}
    \includegraphics[width=0.66\textwidth]{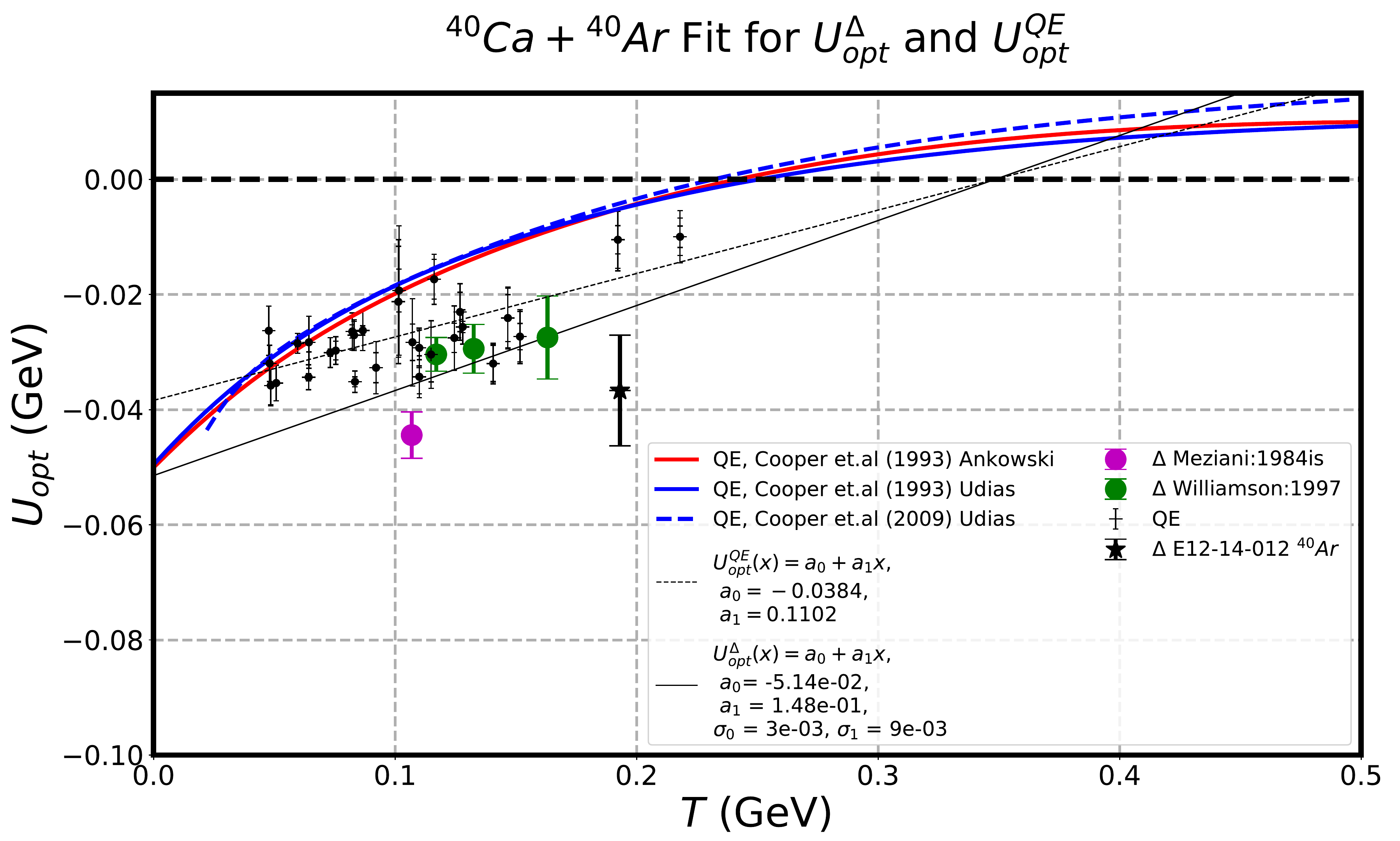}
       \vspace{-0.0cm}
\caption{
\footnotesize\addtolength{\baselineskip}{-1\baselineskip} 
Same as Fig. \ref{UC12} for  $\bf_{20}^{40}Ca$+$\bf_{18}^{40}Ar$.  The top and bottom panels show the measurements  versus $p_{f3}^2= (\vec k +\vec{q}_3)^2$, and versus hadron kinetic energy T, respectively. }
  \label{UCa}
\end{figure*}
%
   For  electron scattering from bound protons,  $V_{eff}$ at the interaction vertex for a  final state proton (in QE scattering), final state  $\Delta^+{1232}$ (in resonance production),  and final state of mass $W^+$ (in inelastic scattering) are defined below. 
    $$ |V_{eff}^P|=  |V_{eff}^{\Delta +}| = |V_{eff}^{W+}| = \frac {Z-1}{Z} |V_{eff}|$$
    For electron scattering from a neutron target we set ${|V_{eff}^N}|=0$.   The values of $|V_{eff}|$ that we use for various nuclei are given in Table \ref{Table1}.
    %
 %
    \section {Removal energy  of initial state nucleons in a nucleus}
   %
 In our analysis we use the impulse approximation.  The nucleon is moving in the mean field (MF) of all  the other nucleons in the nucleus.  The on-shell recoil excited   $[A-1]^*$ spectator nucleus has a momentum  $\vec p_{ (A-1) *}=-\vec k$ and  a mean excitation energy  $\langle {E_x^{P,N}} \rangle$. The off-shell energy of the interacting nucleon is
     \begin{eqnarray}
        \label{eq2} 
   E_i & =&  M_A - \sqrt{ (M_{A-1}*) ^2+\vec k^2} \\
   &=&  M_A - \sqrt{ (M_{A-1}+{{E_x^{P,N}}})^2+\vec k^2} \nonumber \\
   &=& M_{P,N} -\epsilon^{P,N}\nonumber \\ 
    \epsilon^{P,N}& =& S^{P,N} +\langle E_x^{P,N}  \rangle+\frac{\vec k^2}{2M^*_{A-1}}.\nonumber
     \end{eqnarray}
   Here,  $M_P$ = 0.938272 GeV is the mass of the proton,  $M_N$= 0.939565 GeV is the mass of the neutron, and
    $S^{P,N}$ the separation energy (obtained from mass differences of the initial and final state nuclei)  needed to separate the nucleon from the nucleus.  In Ref.\cite{optpaper} we extract the mean excitation energy  $\langle {E_x^{P,N}} \rangle$ (or equivalently  the removal energy $\epsilon^{P,N}$)  using spectral functions measured in  exclusive electron scattering experiments on nuclear targets in which both the final state electron and proton are detected ($ee^\prime P$)
Some of the   neutrino MC generators (e.g. current version of  $\textsc{genie}$\cite{genie}) do not include the effect of the excitation of the spectator nucleus, nor do they include the effects  of the interaction of the final state nucleons and hadrons with the
Coulomb\cite{veff} and nuclear optical potentials of the nucleus.
%
   \section {Average nuclear optical potential for final state nucleons in QE scattering}
 %
We  model the effect of the interaction of final state nucleons with the real part of the  nuclear optical potential with a  parameter  
  $U^{QE}_{opt}(\vec p_{f3}^2)$,  where  $\vec p_{f3}^2$ is the square of the 3-momentum of the final state nucleon at the vertex.  Alternatively, we also extract  $U^{QE}_{opt}(T)$ where $T$ is the kinetic energy of the final state nucleon.  In the analysis we make the assumption that $U^{QE}_{opt}$ for the proton and neutron are the same.
  The parameter $U^{QE}_{opt}(T)$  takes into account on average the effect of the real part of the nuclear optical potential and results in a modification of the energies of both the final state lepton and the final state nucleon.

 The imaginary part of the optical potential results in interactions of the final state nucleon with nucleons in the spectator nucleus. These interactions  can result in the knockout of additional nucleons as well as pion production.  The effect of the imaginary part of the nuclear optical potential (sometimes referred to as Final State Interaction - FSI) is included in current Monte Carlo generators using different models, including cascade models based on measured nucleon-nucleus scattering data.   In this communication we only address the effects of the real part of the nuclear optical potential which modifies the energies of the final state lepton and nucleon. Note that when we refer to the {\it optical potential}, it is a shorthand for the  {\it real part of the optical potential}.
 \begin{figure*}
\centering
  \includegraphics[width=8.cm,height=5.cm]{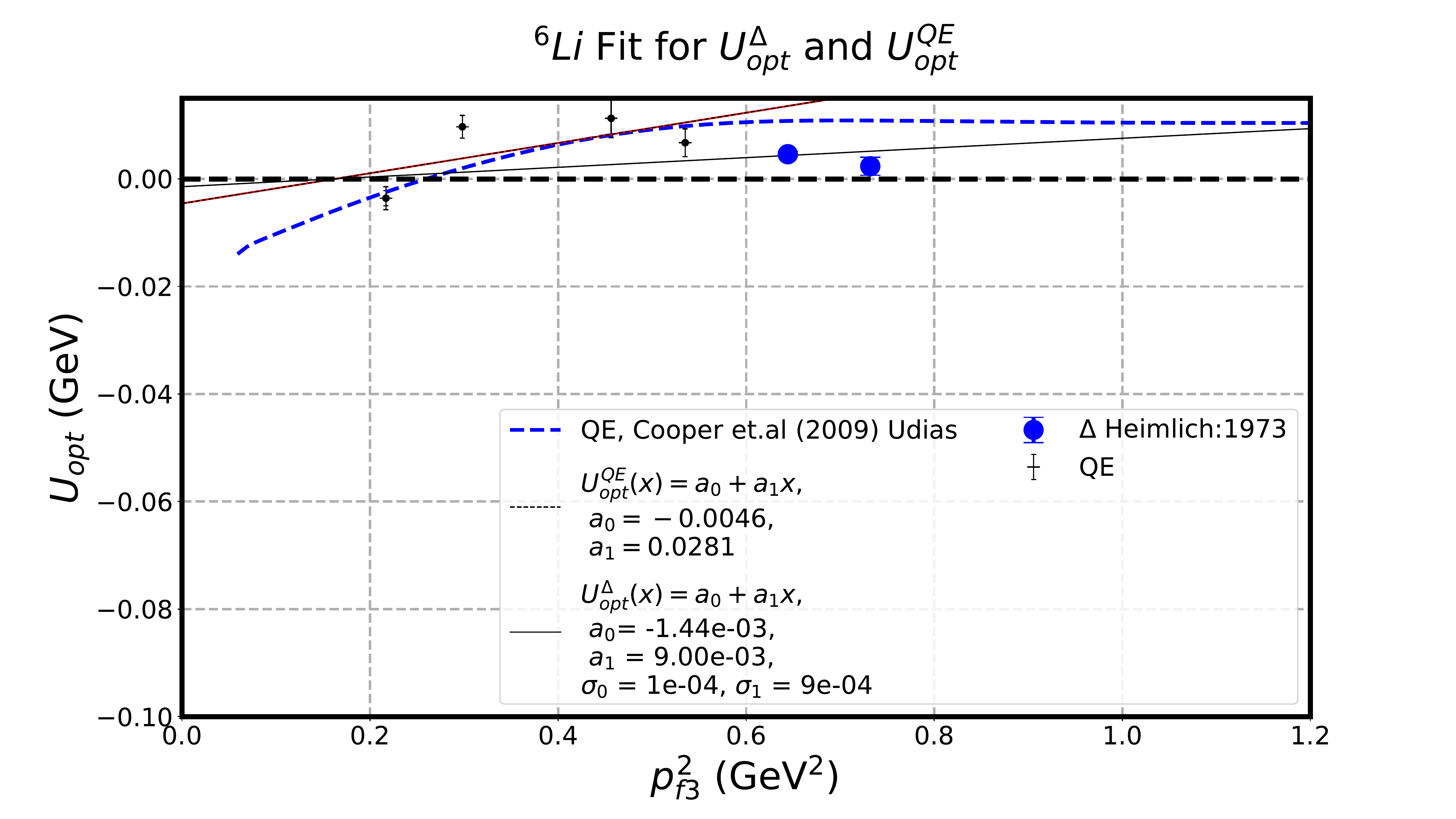}
       \includegraphics[width=8.cm,height=5.6cm]{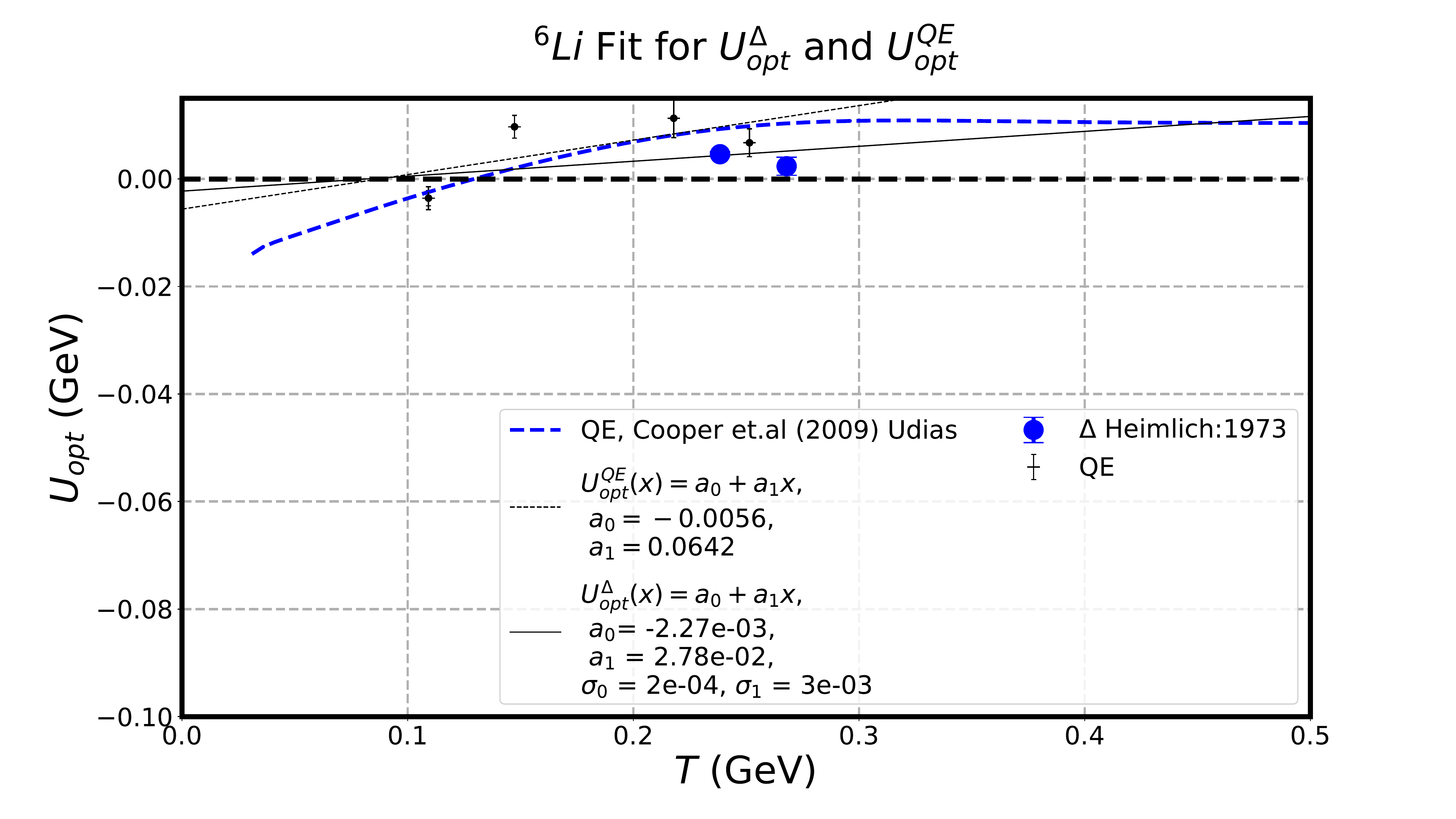}
         \includegraphics[width=8.cm,height=5.6cm]{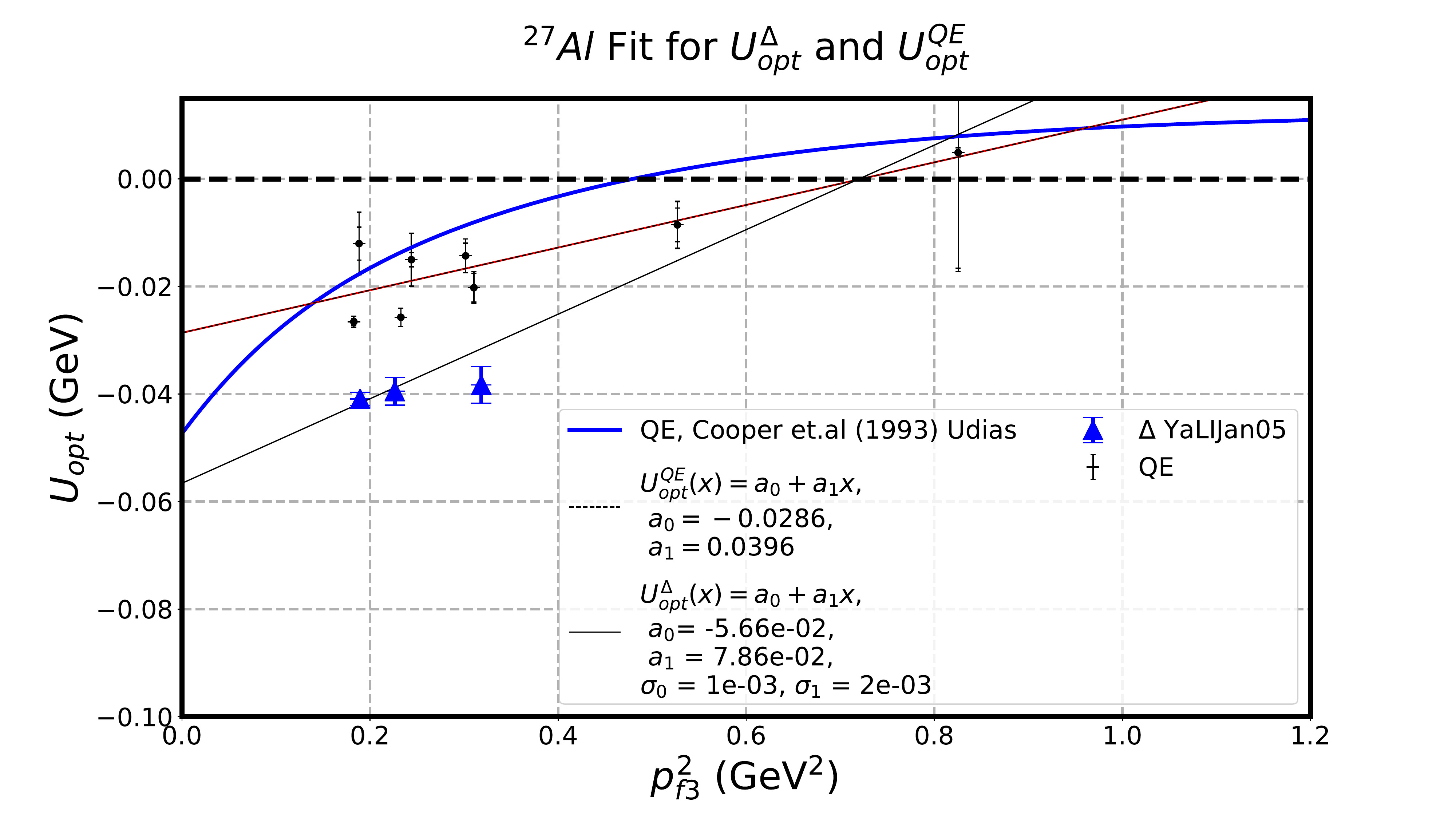}
       \includegraphics[width=8.cm,height=5.6cm]{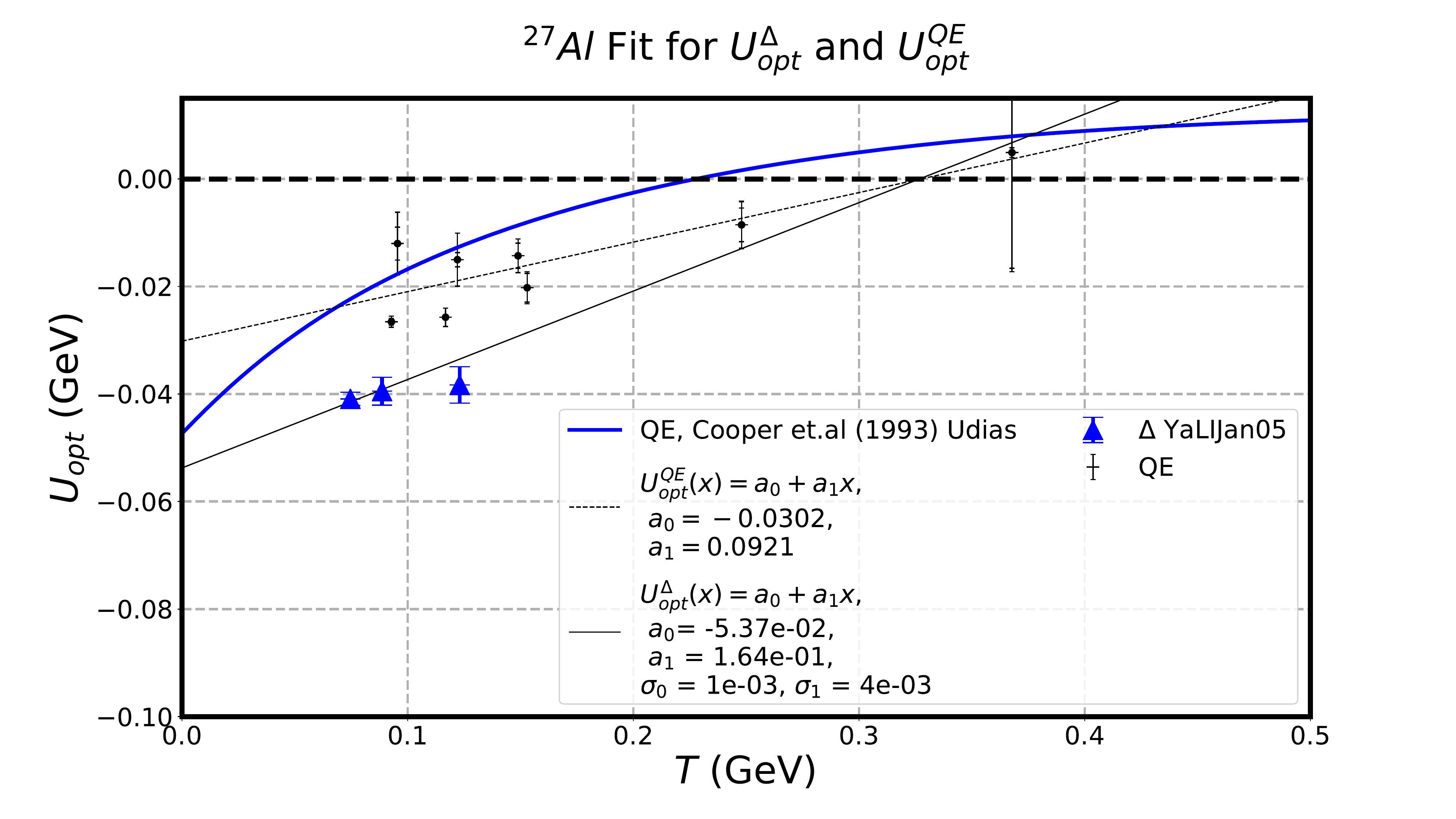}
        \includegraphics[width=8.cm,height=5.6cm]{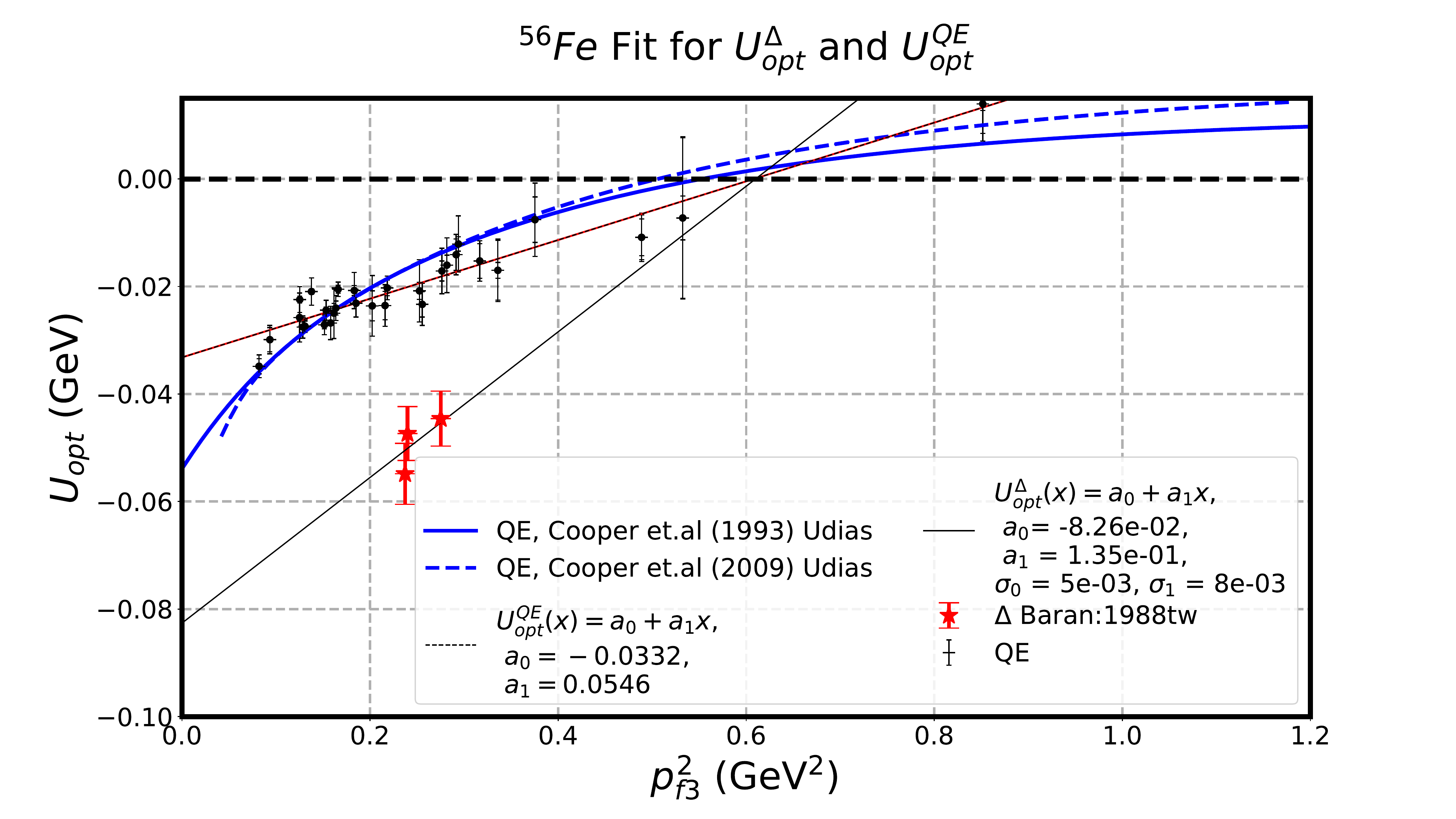}
       \includegraphics[width=8.cm,height=5.6cm]{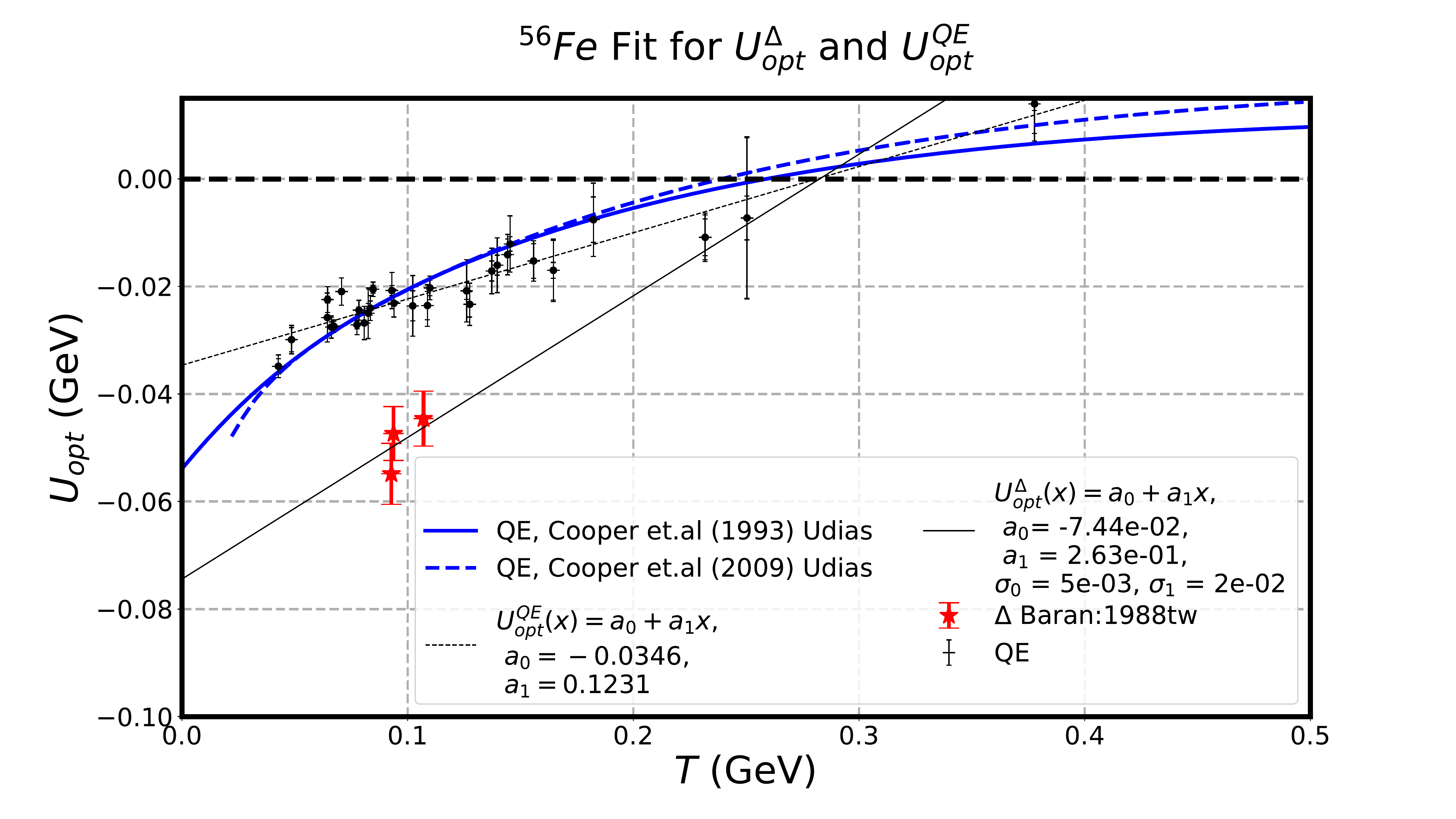}
        \includegraphics[width=8.cm,height=5.6cm]{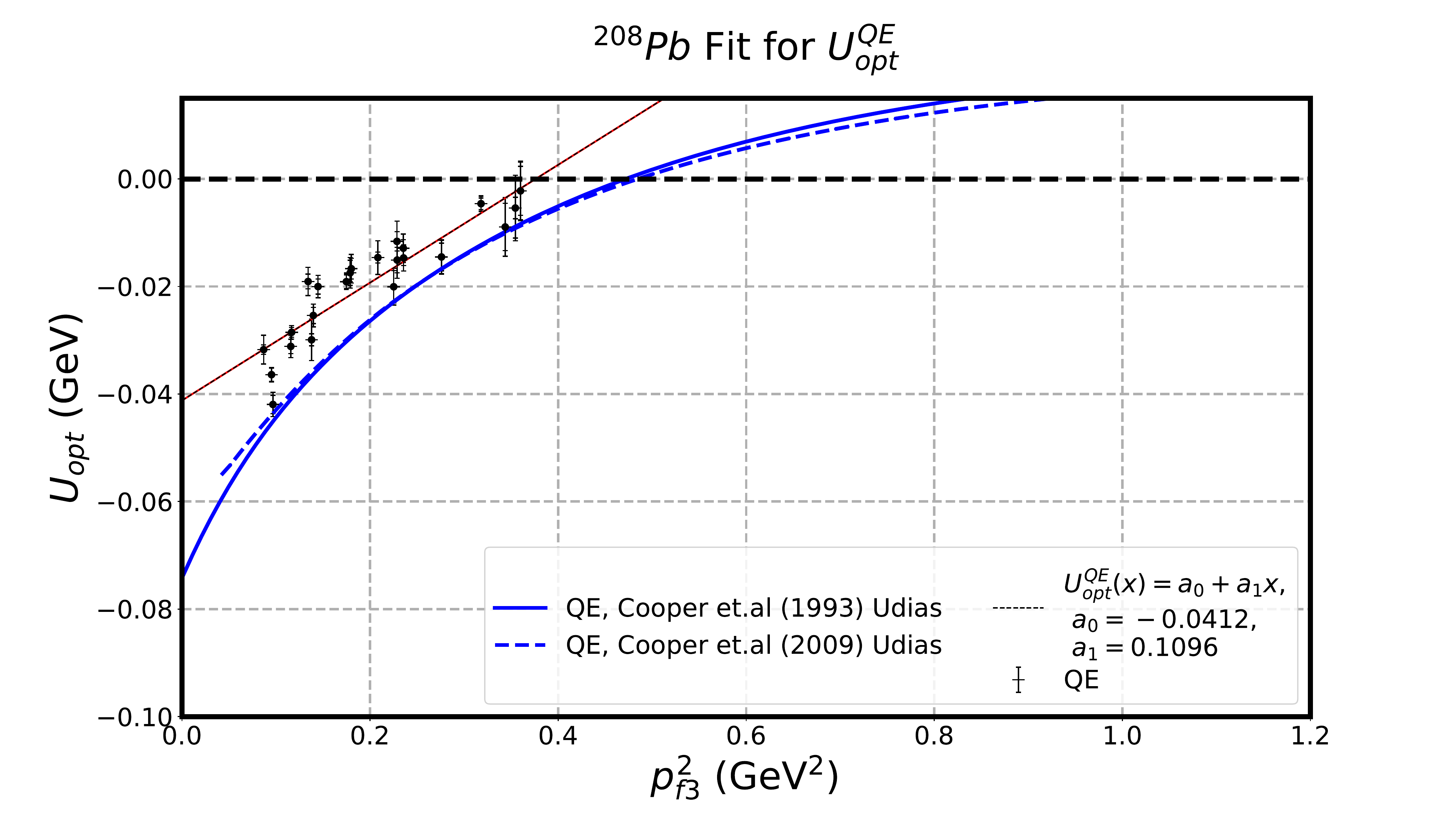}
       \includegraphics[width=8.cm,height=5.6cm]{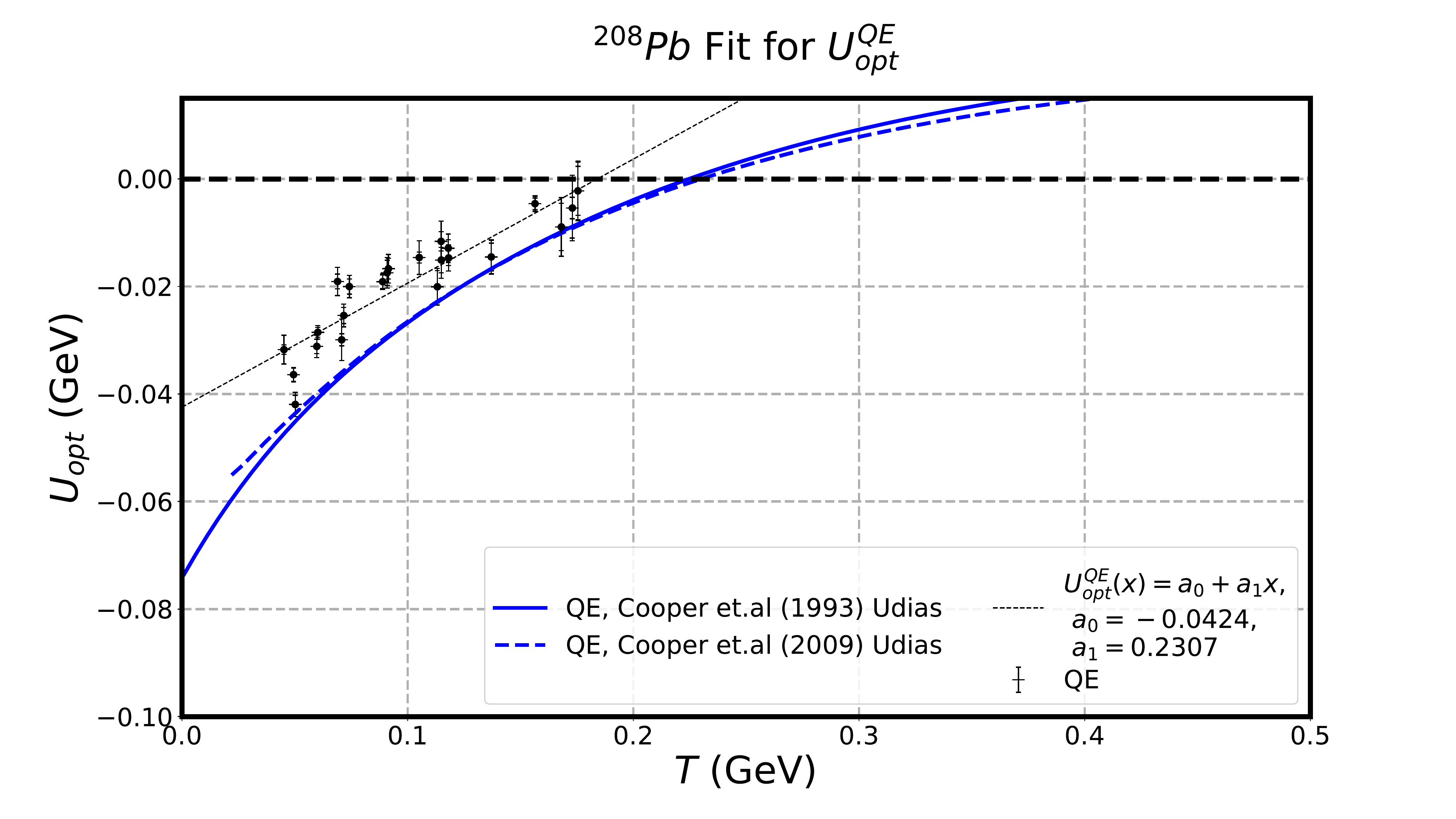}
       \vspace{-0.0cm}
\caption{
\footnotesize\addtolength{\baselineskip}{-1\baselineskip} 
Same as Fig. \ref{UC12} for $\bf_3^{6}Li$, $\bf_{13}^{27}Al$,  $\bf_{26}^{56}Fe$ and $\bf_{82}^{208}Pb$.}
  \label{remaining}
\end{figure*}
 The  energy of the final state nucleon in QE  electron scattering is given by the following expressions:
    \begin{eqnarray} 
    \nu&+&(M_{P,N}-\epsilon^{P,N})= E_f^{P,N}\nonumber \\
    \vec p_{f3}&=&(\vec k +\vec q_3) \nonumber \\
 E_f^{P,N}&=&\sqrt{\vec p_{f3}^2+M_{P,N}^2} +U^{QE}_{opt}(\vec p_{f3}^2)+|V_{eff}^{P,N}|.\nonumber \\
  \label{QE_equation}
 T^{P,N} &=& E_f^{P,N} - M_{P,N},
    \end{eqnarray}
where $T^{P,N}$ is in the kinetic energy of the nucleon of mass $M_{P,N}$ after it leaves the nucleus and is in the same direction as $\vec{p_{f3}}$.
We  extract $U^{QE}_{opt}(\vec p_{f3}^2)$  and  $U^{QE}_{opt}(T)$ from a comparison of the relativistic Fermi gas (RFG) model to measurements of inclusive QE e-A differential cross sections compiled in references  \cite{archive} and \cite{archive1}. 
      The data samples (see references \cite{Heimlich:1974}-\cite{Zghiche:1993xg}) include the following elements which are of interest to current neutrino experiments:  33  $\bf_{6}^{12}C$ spectra, five $\bf_{8}^{16}O$ spectra,  29  $\bf_{20}^{40}Ca$ spectra, and  two  $\bf_{18}^{40}Ar$ spectra.  
       
 In addition,  the data sample include four  $\bf_{3}^{6}Li$ spectra, eight $\bf_{18}^{27}Al~$spectra,   30 $\bf_{26}^{56}Fe$  spectra,  and  23 $\bf_{82}^{208}Pb$~spectra.
 Most (but not all) of the QE differential cross sections  are available on the  QE electron scattering archive\cite{archive,archive1}.  
      
 In the extraction of the average  nuclear optical potential for final state nucleons in QE scattering we only fit to the data in the top 1/3 of the QE distribution and extract  the best value of $U^{QE}_{opt} (\vec p_{f3}^2)$ and $U^{QE}_{opt} (T)$. Here  $\vec p_{f3}$ is evaluated at the peak of the QE distribution.  In the fit we let the normalization of the QE cross section float to agree with data. 
 Figure \ref{C12_fits} shows examples of two of the 33 fits to QE  differential cross sections for $\bf_{6}^{12}C$. The solid black curves are the  RFG fits with the best value of $U^{QE}_{opt}$ for the final state nucleon.
   The blue dashed curves are simple parabolic fits used to estimate the systematic error.  The difference between $\nu_{peak}^{parabola}$ and $\nu_{peak}^{rfg}$ is used  as a systematic error in our  extraction of $U^{QE}_{opt}$.   The first error shown in the legend of  Figure \ref{C12_fits} is the statistical error in the fit.  The second error is the systematic uncertainty, which is much larger.  The red dashed curve is the RFG model  with  $U^{QE}_{opt}=0$ and $|V^P_{eff}|=0$.
  
The extracted  values of  $U^{QE}_{opt}(\vec p_{f3}^2)$ versus $\vec p_{f3}^2$ from  33  $\bf_{6}^{12}C$ QE spectra and five $\bf_{8}^{16}O$ QE spectra are shown in the top panel of  Figure \ref{UC12}.   The same values as a function of the nucleon kinetic energy T are shown on the bottom panel.   The extracted values of   $U^{QE}_{opt}(\vec p_{f3}^2)$ versus $\vec p_{f3}^2$  (and T)  from   29  $\bf_{20}^{40}Ca$ QE  spectra and two $\bf_{18}^{40}Ar$  QE spectra  are shown in Fig. \ref{UCa}.   

Note that the figures also show values of the average  optical potential for the $\Delta$ resonance which is discussed in a later section of this paper.

Similarly, values extracted  for four $\bf_{3}^{6}Li$ QE spectra, eight  $\bf_{18}^{27}Al$ QE spectra,  30 $\bf_{26}^{56}Fe$  QE spectra and   23 $\bf_{82}^{208}Pb$ 
 QE spectra are shown in Fig.\ref{remaining}.
 
  We fit the extracted values of $U^{QE}_{opt}(\vec p_{f3}^2)$ versus $\vec p_{f3}^2$ for $\vec p_{f3}^2>0.1$ GeV$^2$ to linear functions which are shown as as dashed grey lines in Figures \ref{UC12}-\ref{remaining}.  We also show linear fits to  $U^{QE}_{opt}$ as a function of final state kinetic energy T.
  The intercepts  at $\vec p_{f3}^2=0$ and the slopes of the fits to  $U^{QE}_{opt}$ versus $\vec p_{f3}^2$, and the intercepts and slopes of the fits to  $U^{QE}_{opt}$ as a function of T  are given in Table \ref{Table1}. Fits for $\bf_{6}^{12}C$ using a different functional form are give in Table  \ref{Table2} of the Appendix.
  
    Note that parameters for the average optical potential for the $\Delta$ resonance (discussed in a later section of this paper) are also included in  Tables \ref{Table1} and in Table \ref{Table2}.
  \subsection {Comparison of the values of $U^{QE}_{opt}$ to calculations} 
  %
The formalism of the nuclear optical potential of  Cooper, Hama, Clark and Mercer\cite{Cooper1993,Cooper2009} is phenomenological. They propose a few parametrizations of the optical potential, and determine their dependence on the kinetic energy of the nucleon and radial coordinate by fitting the scattering solutions to the proton-nucleus data for the elastic cross sections, analyzing powers, and spin rotation functions for proton scattering on different nuclei.  Since electron and neutrino interactions can occur at any location in the nucleus, it is the  average value of the optical potential $U^{QE}_{opt}$ over the entire nucleus that is the parameter that is needed for MC simulations.

The solid blue lines in  Figures \ref{UC12},  \ref{UCa},  and  \ref{remaining} are    the average nuclear optical potential for final state nucleons $U^{QE}_{opt}$ calculated by  Jose Manuel Udias\cite{Udias}  using the formalism of Cooper, Hama,  Clark  and Mercer published in 1993\cite{Cooper1993}, and the  dashed blue lines are  calculated using the later formalism of Cooper, Hama and Clark published in 2009\cite{Cooper2009}.

As a check, the solid red lines in  in  Figures \ref{UC12} and  \ref{UCa} are  calculations of 
$U^{QE}_{opt}$ by  Artur. M. Ankowski\cite{Artur} using Cooper 1993\cite{Cooper1993} formalism, and the dashed red lines are calculated using the Cooper  2009\cite{Cooper2009} formalism.

The measurements of  $U^{QE}_{opt}$  for $\bf_{3}^{6}Li$ and $\bf_{26}^{56}Fe$ are in good  agreement with the  Cooper 1993\cite{Cooper1993}  and Cooper 2009\cite{Cooper2009} calculations. The measurements are less negative than the calculations for $\bf_{82}^{208}Pb$.
 The measurements are more negative than the calculations for $\bf_{6}^{12}C$+$\bf_{8}^{16}O$, $\bf_{18}^{27}Al$, and  $\bf_{20}^{40}Ca$+$\bf_{18}^{40}Ar$.  For the $\bf_{6}^{12}C$ nucleus, although both   calculations  of  $U^{QE}_{opt}$ are above the data, the Cooper 1993\cite{Cooper1993} calculations are closer to the data than the Cooper 2009\cite{Cooper2009} calculations.   
 \begin{figure} 
\centering
    \includegraphics[width=3.3in,height=2.6in]{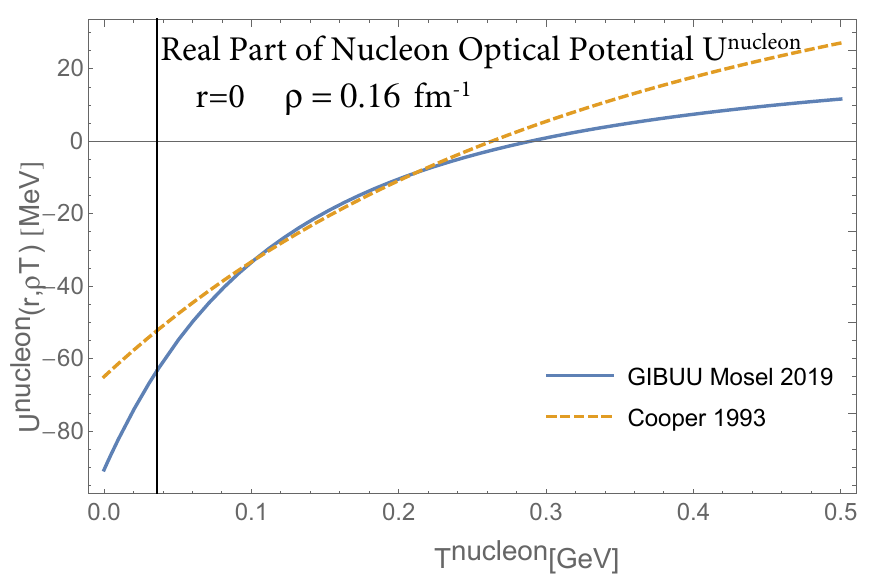}
  \caption{
\footnotesize\addtolength{\baselineskip}{-1\baselineskip} 
A comparison of the real part of the nucleon optical potential
 ($U^{nucleon}(r,\rho,T)$) 
 versus kinetic energy $T$ at r=0 and nuclear density $\rho$= 0.16 fm$^{-1}$ for  GiBUU 2019\cite{GiBUU} as compared to the potential parametrized by  Cooper 1993\cite{Cooper1993} (curves from Ref. \cite{GiBUU}). The two optical potentials are  consistent with each other between nucleon kinetic energy between 0.1 and 0.3 GeV$^2$ (curves from Ref. \cite{GiBUU}).}
\label{GiBUU_Optical}
\end{figure} 
\subsection{Discussion of the optical potential for nucleons}
%
 We  extract the parameter $U^{QE}_{opt}$ which the average of the real part of the optical potential for the final state nucleon.  This parameter is extracted for use in Monte Carlo generators for which the initial state nucleon is described by a spectral function.  The spectral function describes the momentum distribution of the nucleon and its removal energy and can be measured in exclusive  $ee^\prime P$  electron scattering experiments on  nuclear targets,   Since the nucleon can be removed from any location in the nucleus, it is the average optical potential that is the relevant parameter.  Since we use the measured removal energies to describe the initial state nucleon, the potential which binds the nucleon  is not relevant to our analysis.

The GiBUU \cite{GiBUU} model describes the initial state as a bound nucleon for which the momentum distribution is related to the local density of nucleons  $\rho$. The nuclear  potential  $U(\rho, T)$ for the initial state nucleon depends on both the local density and nucleon momentum.  The same density and kinetic energy dependent  potential is used for the initial and final state nucleons.
  Figure \ref{GiBUU_Optical} shows
 a comparison of the real part of the nucleon optical potential
 ($U^{nucleon}(r,\rho,T)$) 
 versus kinetic energy $T$ at r=0 and nuclear density $\rho$= 0.16 fm$^{-1}$ for  GiBUU 2019\cite{GiBUU} as compared to the potential parametrized by  Cooper 1993\cite{Cooper1993} (curves from Ref. \cite{GiBUU}).  The two optical potentials are  consistent with each other for nucleon kinetic energies between  0.1  and 0.3 GeV$^2$. 
  \section {Average nuclear optical potential for a $\Delta$ resonance in the final state} 
%
 %
Several theoretical groups model the quasielastic and $\Delta$ production in nuclear targets.   The Valencia group\cite{Valencia} uses a local Fermi gas model with RPA correlations. The model accounts for medium effects through the use of nucleon-hole and $\Delta$-hole spectral functions.  The Giessen group uses the GiBUU\cite{GiBUU, GiBUU2} implementation of quantum-kinetic transport theory to describe the QE and $\Delta$ regions.  As mentioned earlier an ingredient in GiBUU is a momentum dependent potential translated into an effective nucleon mass. A summary of various models can be found in reference\cite{white_paper}.
 
\begin{figure*}
\begin{center}
\includegraphics[width=3.5in,height=2.6in]{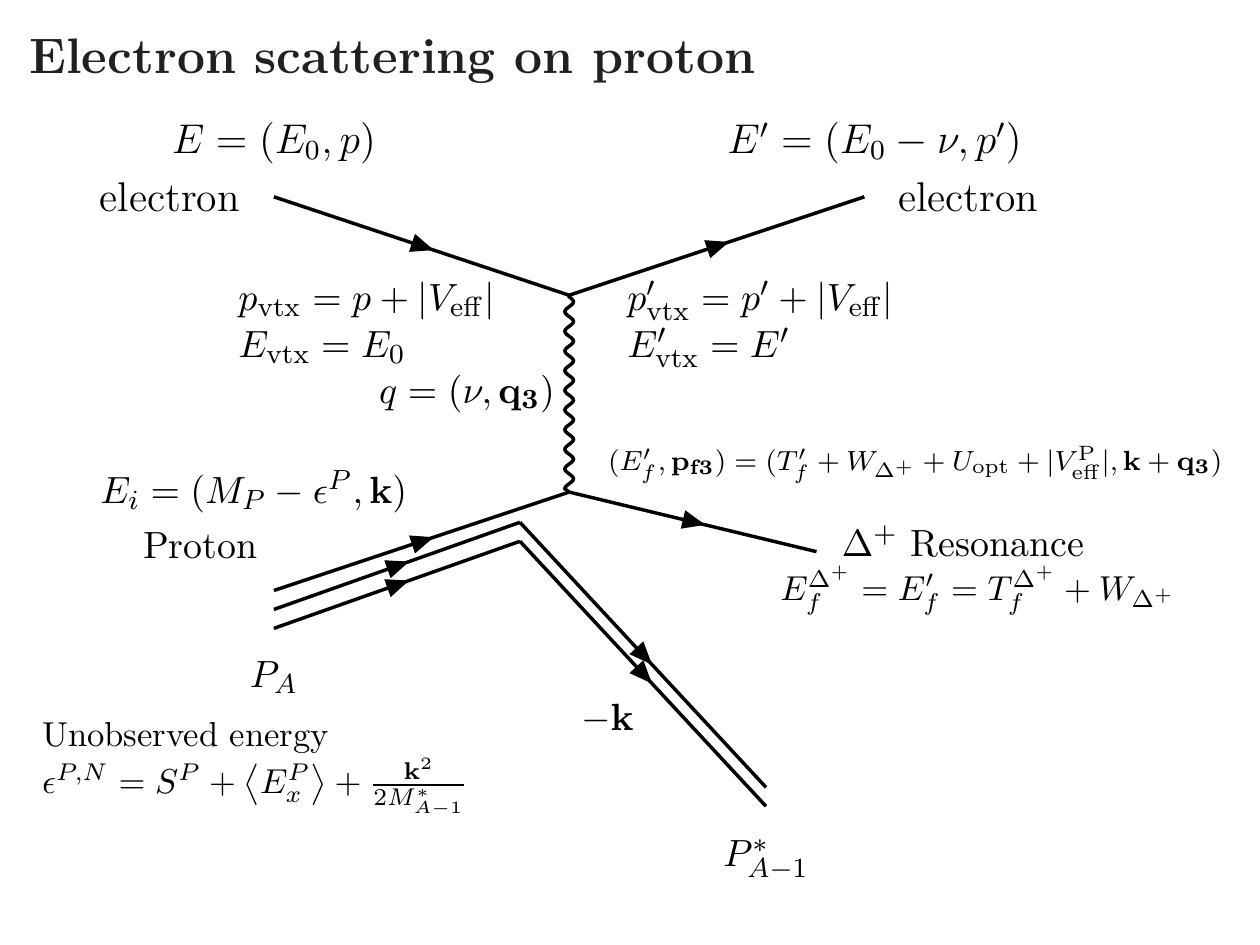}
\includegraphics[width=3.5in,height=2.6in]{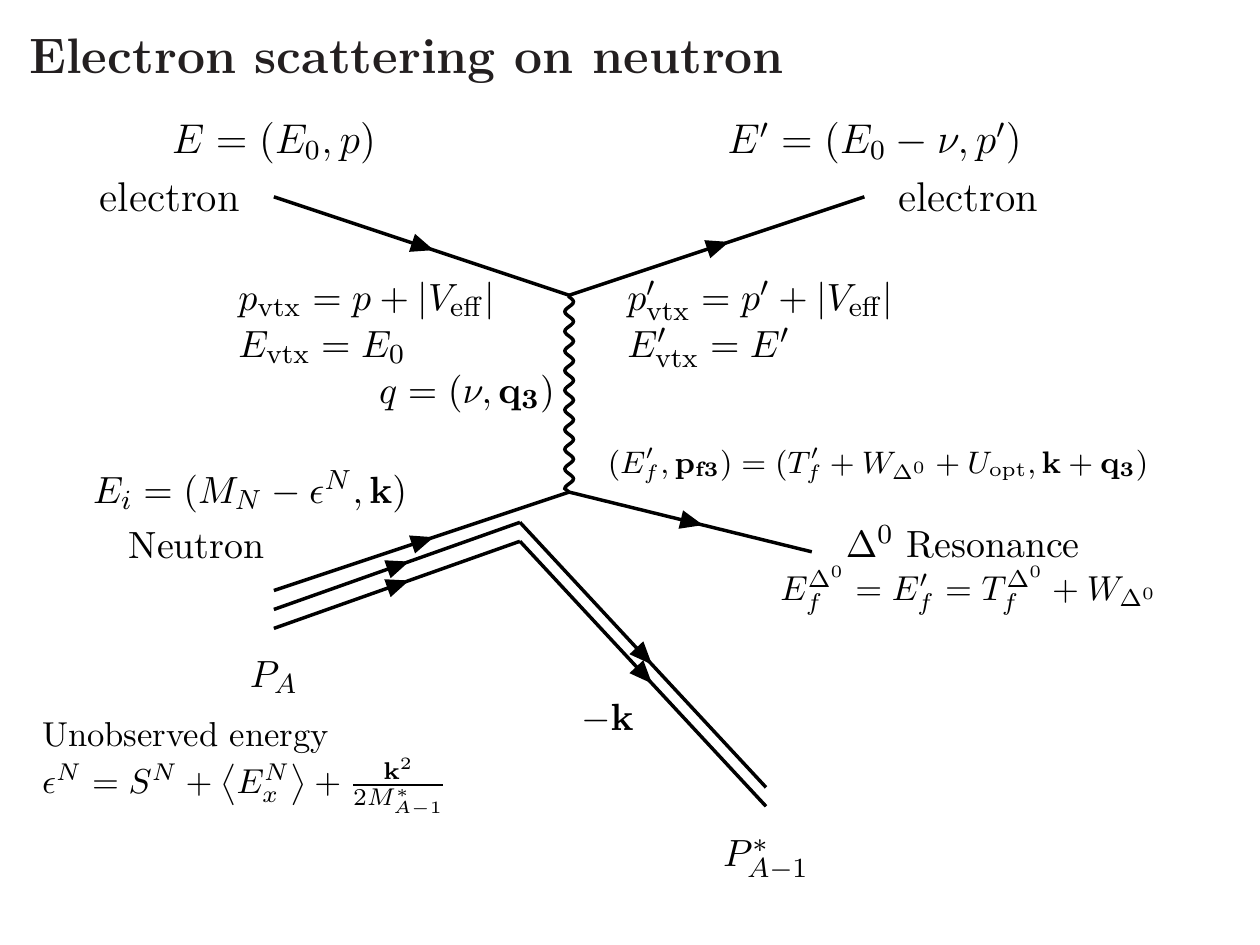}
\includegraphics[width=3.5in,height=2.6in]{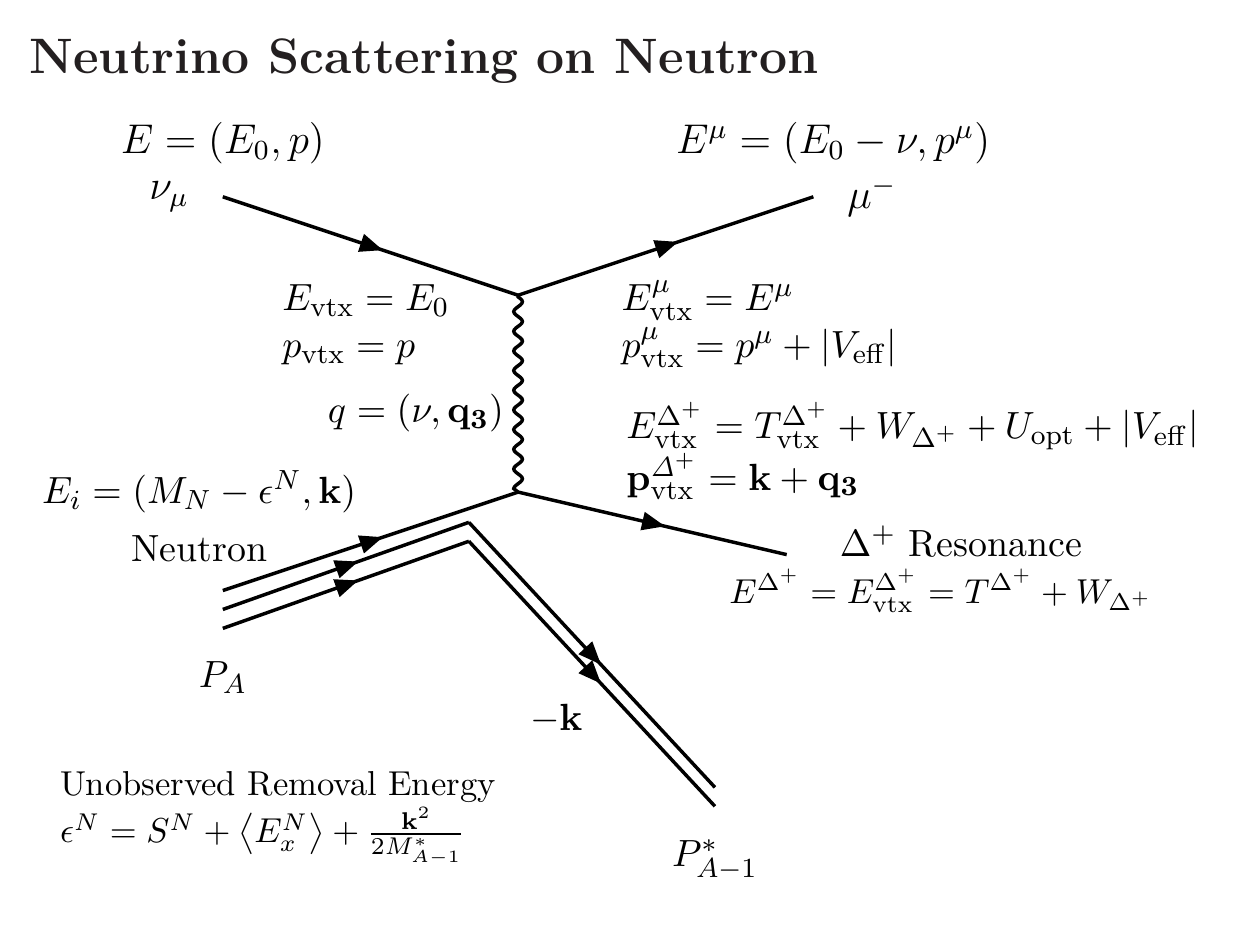}
\includegraphics[width=3.5in,height=2.6in]{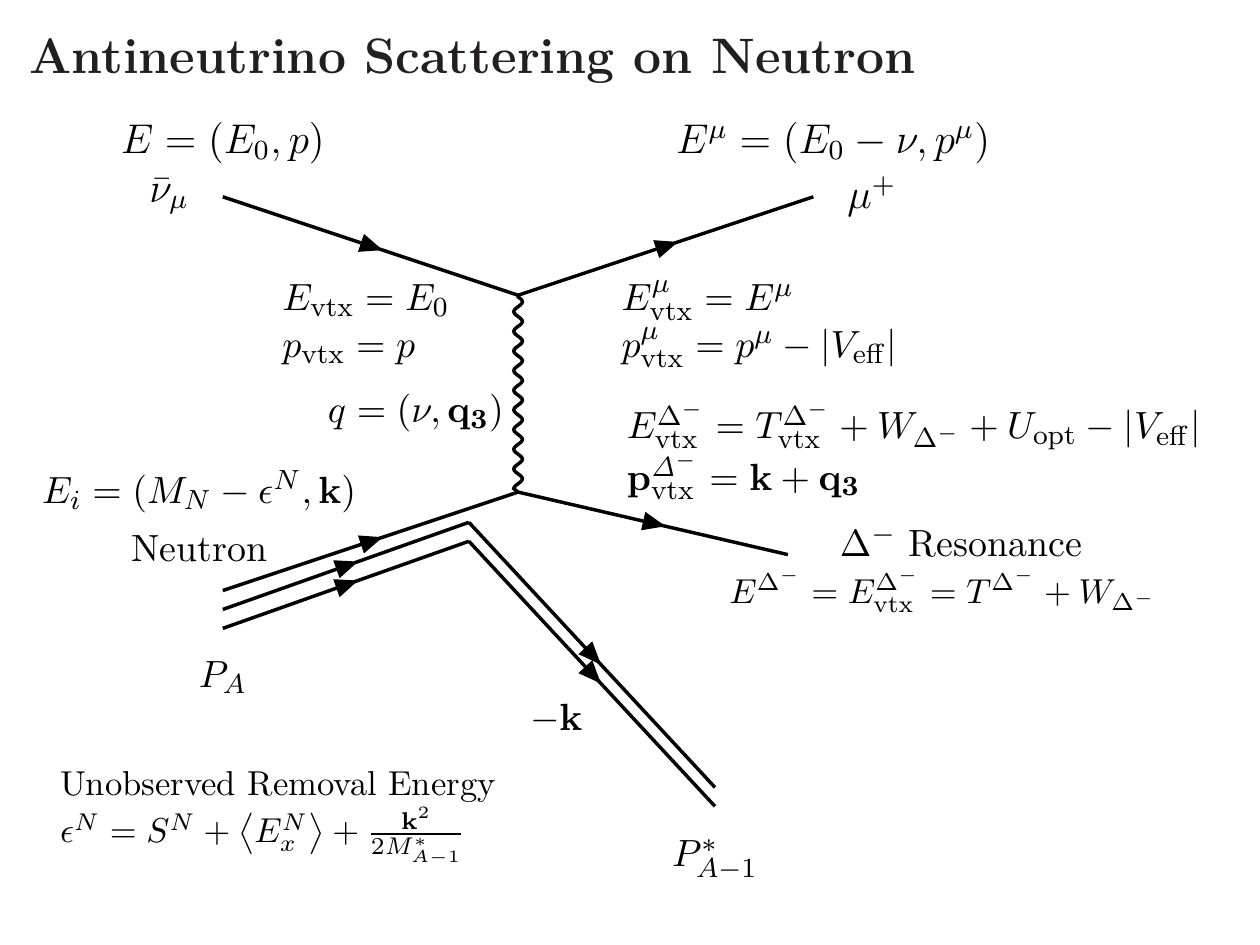}
%
\caption{
The top two panels show diagrams for electron scattering from a bound proton producing an invariant mass W in the region of the $\Delta^{+}$(left), and scattering from a bound neutron producing an invariant mass W in the region of the  $\Delta^{0}$(right).  The bottom two panels show neutrino scattering from a bound neutron producing an invariant mass W in the region of the   $\Delta^{+}$(left) and antineutrino scattering on a bound neutron producing an invariant mass W in the region of the  $\Delta^{-}$(right).}
\label{e_Delta}
\end{center}
\end{figure*}

The top two panels in Fig. \ref{e_Delta} show diagrams for electron scattering from a bound proton producing an invariant mass W in the region of the $\Delta^{+}$(left), and scattering from a bound neutron producing an invariant mass W in the region of the  $\Delta^{0}$(right).  The bottom two panels show neutrino scattering from a bound neutron producing an invariant mass W in the region of the   $\Delta^{+}$(left) and antineutrino scattering on a bound neutron producing an invariant mass W in the region of the  $\Delta^{-}$(right).

For electron scattering from a bound nucleon the average optical potentials for QE electron scattering and the production of an invariant mass W in the region of the $\Delta$ resonance are defined as follows: 
 \begin{eqnarray}
 \label{Delta_Eq}
  \nu &+&(M_{P,N}-\epsilon^{P,N}) =E_f  \\
  E_f^{P} &= &\sqrt{ (\vec k +\vec q_3)^2+M_{P}^2} +U^{QE}_{opt}+  |V_{eff}^P| \nonumber\\
  E_f^{N} &= &\sqrt{ (\vec k +\vec q_3)^2+M_{N}^2} +U^{QE}_{opt}  \nonumber\\
   T^{P,N} &=& E_f^{P,N} - M_{P,N}\nonumber \\
   E_f^{\Delta+}&=& \sqrt{ (\vec k +\vec q_3)^2+W_{\Delta+}^2} +U^{\Delta}_{opt}+  |V_{eff}^{\Delta+}| \nonumber \\
     E_f^{\Delta0}&=& \sqrt{ (\vec k +\vec q_3)^2+W_{\Delta0}^2} +U^{\Delta}_{opt}\nonumber\\ 
     T^{\Delta(+,0)}&=& E_f^{\Delta(+,0)}-W_{\Delta(+,0)},\nonumber
   \end{eqnarray}
where $W_{\Delta+,0}$  is the final state invariant  mass in the region of the  $\Delta$ resonance and   $|V_{eff}^{\Delta+}|=|V_{eff}^P|=\frac{Z-1}{Z} |V_{eff}|$.   Here $T^{\Delta(+,0)}$ is the kinetic energy of the resonance of mass $W_{\Delta(+,0)}$ after it leaves the nucleus and is in the same direction as $\vec{p_{f3}}$.
  %
  %
%
   \begin{figure*}
\centering
 \includegraphics[width=0.42\textwidth]{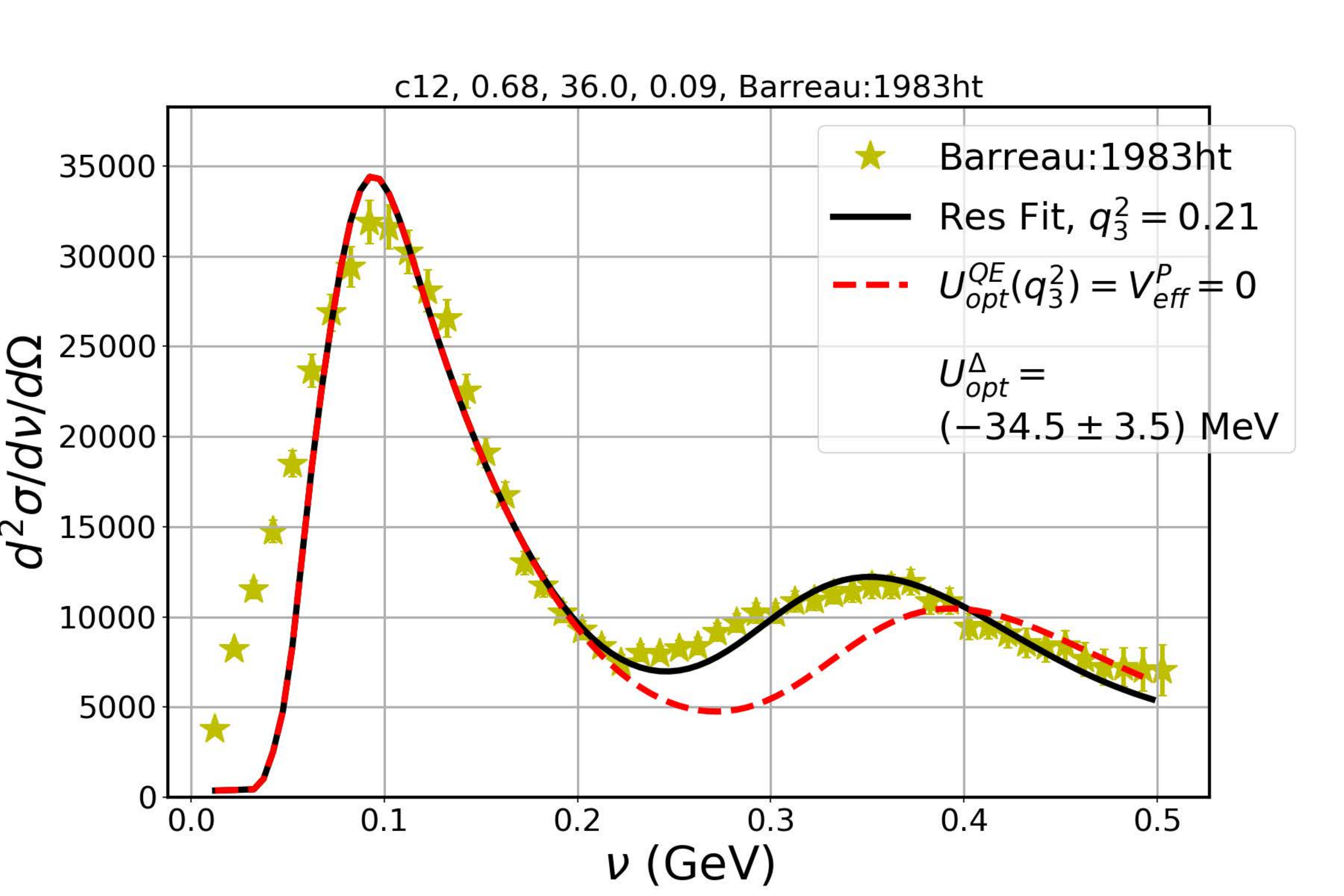}
\includegraphics[width=0.42\textwidth]{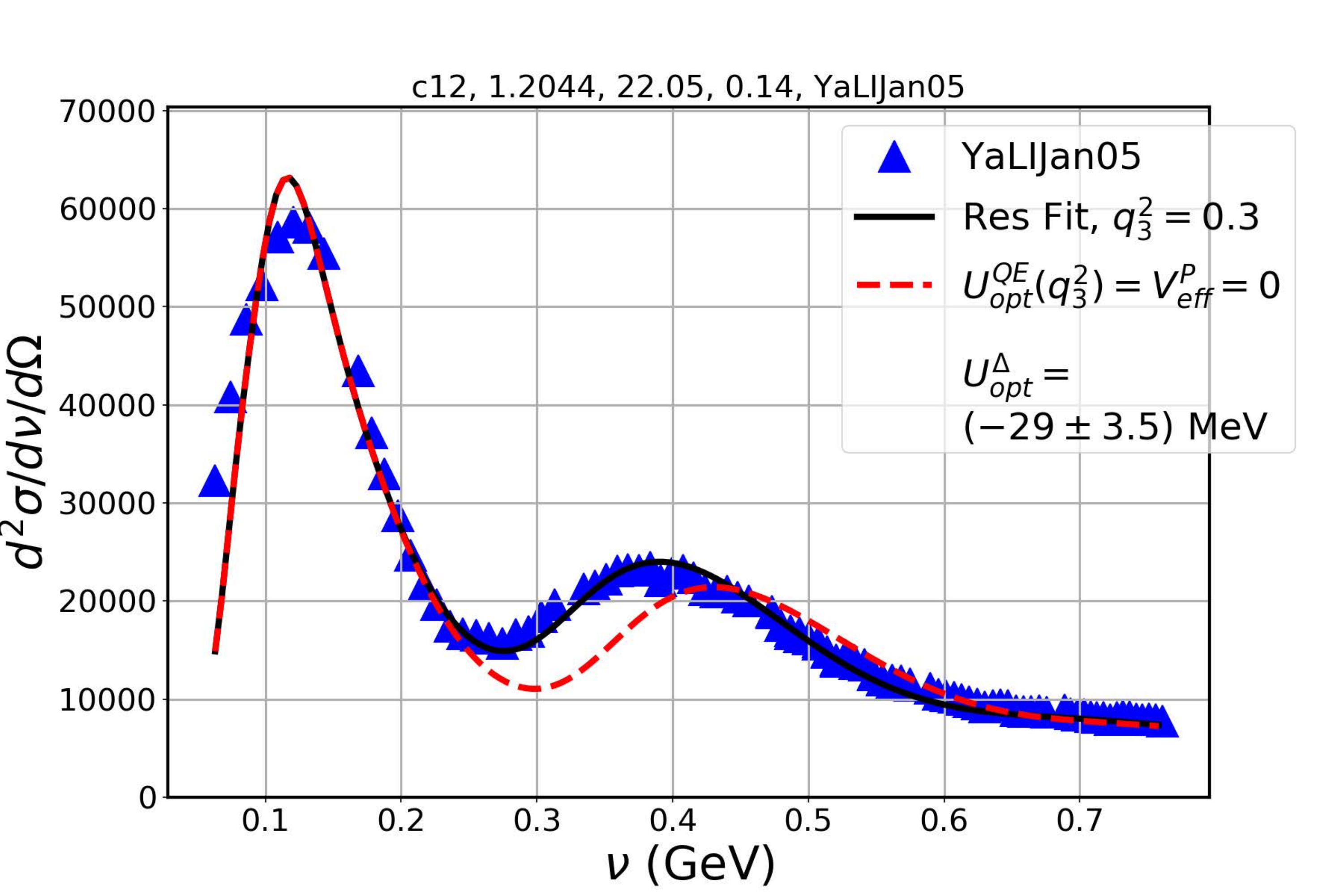}
\caption{
\footnotesize\addtolength{\baselineskip}{-1\baselineskip} 
Examples of fits for  two out of  15  $\bf_{6}^{12}C$ $\Delta$(1232) production differential cross sections. Here the QE peak is modeled with an effective spectral function (including 2p2h), and $\Delta$ production is modeled by using RFG to smear fits to resonance production (and continuum) structure functions on free nucleons. The solid black curves are the fit with the best value of $U^\Delta_{opt}$. The  dashed red  curves are the predictions  with  $U^\Delta_{opt}=V^\Delta_{eff}=0$.
}
\label{C12_delta_fits}
\end{figure*}
%
%
   %
 \begin{table*}
 \begin{center}					
\begin{tabular}{|ccc|cc||cc|}					
\hline 					
$\bf_Z^{A}Nucl$	& $|V_{eff}|$  & removal &$U_{opt}$  & $U_{opt}$ & $U_{opt}$  & $U_{opt}$ 
        \\ 
	& MeV& energy&	intercept 	& slope vs &	intercept 	& slope vs
	 \\
	&&$\epsilon^{P},\epsilon^N$&	$\vec p_{f3}^2$	& $\vec p_{f3}^2$ & $T$=0	& $T $  \\ 
	& & MeV&	$=(\vec q_3+\vec k)^2$=0	& $=(\vec q_3+\vec k)^2$ &GeV&GeV/GeV$^2$   \\ 
\hline \hline
$_3^{6}Li (QE~proton)$ &1.4& 18.4,19.7
 &-0.005$\pm$0.010 & 0.028$\pm$0.028 
&-0.006$\pm$0.011 & 0.064$\pm$0.057    \\  
$_3^{6}Li~\Delta^+ / \Delta^0 \rightarrow$ & $\pm$0.25& ($\pm$3.0) 
&-0.001$\pm$0.001  &0.009$\pm$0.001  
& -0.002$\pm$0.001  &0.028$\pm$0.003    \\
\hline 
$_6^{12}C + _8^{16}O (QE~proton)*$ &3.1&27.5, 30.1	
&-0.029$\pm$0.004& 0.040$\pm$0.010 
&-0.030$\pm$0.001& 0.092$\pm$0.009   \\ 
$_6^{12}C~\Delta^+ / \Delta^0 \rightarrow$ &$\pm$0.25&($\pm3.0$)    
&-0.049$\pm$0.001  &0.059$\pm$0.001    
& -0.050$\pm$0.001&0.161$\pm$0.001   \\
 \hline 
 $_{13}^{27}Al(QE~proton)$ &5.1&	   30.6, 35.4       
 &-0.029$\pm$0.004& 0.040$\pm$0.010 
 &-0.030$\pm$0.004& 0.092$\pm$0.023  \\ 
$_{13}^{27}Al~\Delta^+ / \Delta^0 \rightarrow$&$\pm$0.6 &($\pm$3.0)
&-0.059$\pm$0.001  & 0.079$\pm$0.002  
&-0.054$\pm$0.001 &0.164$\pm$0.004   \\
 \hline 
$_{20}^{40}Ca+_{18}^{40}Ar(QE~proton)*$ &7.4/6.3&28.2, 35.9 	 
&-0.038$\pm$0.002& 0.052$\pm$0.010 
&-0.038$\pm$0.002& 0.110$\pm$0.023  	\\ 
$_{20}^{40}Ca~\Delta^+ / \Delta^0 \rightarrow$ &$\pm$0.6  &  ($\pm$3.0)
&-0.059$\pm$0.004 &  0.083$\pm$0.005 
& -0.051$\pm$0.003&0.148$\pm$0.009   \\
\hline 
$_{26}^{56}Fe(QE)$ &8.9  &  29.6, 30.6      
&-0.033$\pm$0.001& 0.055$\pm$0.003 
&-0.035$\pm$0.001& 0.123$\pm$0.008  	\\  
$_{26}^{56}Fe~ \Delta^+ / \Delta^0 \rightarrow$&$\pm$0.7& ($\pm$3.0
)&-0.083$\pm$0.005\  &0.135$\pm$0.008 
&-0.074$\pm$0.005 &0.263$\pm$0.020    \\
\hline 
$_{82}^{208}Pb(QE)$ &18.9 &22.8, 25.0 
&-0.041$\pm$0.002 & 0.110$\pm$0.011
&-0.042$\pm$0.003 & 0.231$\pm$0.023  	\\  
&  $\pm$1.5&($\pm$5.0) & &   & &     \\
\hline 
Average all nuclei &  & & &     &-&    \\
$\Delta^+ / \Delta^0 \rightarrow$&  & &-0.062$\pm$0.001 & 0.101$\pm$0.002    &-0.064$\pm$0.001 & 0.284$\pm$0.005   \\
\hline  
\end{tabular}
\caption{ The second column shows values of   $|V_{eff}|$ (MeV) for various nuclei.  The third column shows the removal energies for protons and neutrons (MeV).
 The 4th and 5th columns show
the intercepts (GeV) at $\vec p_{f3}$=0 and slopes (GeV/GeV$^2$)
 of linear  fits to  $U^{QE}_{opt}$ and $U^{\Delta}_{opt}$ versus  $\vec p_{f3}^2=(\vec q_3+\vec k)^2$.  
 The 6th and 7th columns show the results of a similar analysis versus the final state kinetic energy T. The overall  systematic error on  $U^{QE}_{opt}$ is estimated at $\pm$0.005 GeV.  We show the slopes and intercepts for $U^{QE}_{opt}$ and  $U^{\Delta}_{opt}$ on  alternate rows. (*The removal energies\cite{optpaper}  are (24.1,27.0) for $\bf_8^{16}O$ and (30.9, 32.3) for $\bf_{18}^{40}Ar$).  Note that the fits are only valid in the region for which we have data.  Fits for $\bf_{6}^{12}C$ using a different functional form are give in Table \ref{Table2}.
}	
\label{Table1} 					
\end{center}					
\end{table*}

In order to extract the average nuclear  optical potential for a $\Delta$ resonance we need to model the cross section between the QE peak and the $\Delta$ resonance.  We use the effective spectral function\cite{effective} (which includes a 2p2h contribution) to model the region of the  QE peak.    In the calculation of  the inelastic  cross section for the production of resonances and the continuum we use the Bosted-Christy fits\cite{eric} to the inelastic structure functions for free protons and neutrons in the resonance region and continuum. As described in the Appendix,  these are fits a wide range of inelastic electron scattering data on protons and deuterons including photoproduction data at $Q^2=0$. 
The fits   describe the inelastic structure functions for protons and neutrons including both resonances and continuum over a wide range of $Q^2$.
     
   The Bosted-Christy fits to the proton and neutron structure functions are smeared with a  relativistic  Fermi gas (RFG) to model the resonance production from nuclei. When comparing to data, the  $\Delta$ average optical potential is included as a free parameter in the fit. We use a subset of the measured electron scattering cross sections on nuclei that includes measurements of both QE and resonance production. To  extract values of  the average nuclear optical potential for a  $\Delta$(1232) resonance in the final state ($U_{opt}^\Delta$) we compare the data to  predictions of the sum of  QE  and Fermi smeared resonance production cross sections. In the fits the normalizations of the QE cross section,  resonance cross sections  and $U_{opt}^\Delta$ are varied to fit the data. We only include spectra for which the the inelastic continuum is small and the  $\Delta$ resonance can be clearly identified. We do not include high $Q^2$ data  because the Fermi smearing from the continuum and higher mass resonances is  significant and  the uncertainty in the  determination of the peak of the $\Delta$ resonance is much larger.

The systematic error in the extracted value of  $U_{opt}^\Delta$ is obtained by changing  the value of  $U_{opt}^{QE}$  by $\pm$50\%.  This shifts the the location of the QE peak relative to the $\Delta$.  The best values of $U_{opt}^\Delta$ are extracted and the difference in the two extracted values of $U_{opt}^\Delta$  is taken as the systematic error.  
\subsection{Extraction of $U^{\Delta}_{opt}$ from data}
%
Examples of fits for  two out of  15   $\Delta$(1232) production differential cross sections on $\bf_{6}^{12}C$ are shown in
 Fig. \ref{C12_delta_fits}. The solid black curves are the fits with the best value of $U^\Delta_{opt}$. The  dashed red  curves are the same fits  with $U^\Delta_{opt}$ and $|V^\Delta_{eff}|$ set to zero.
The extracted  values of $U^{QE}_{opt}$ and   $U_{opt}^\Delta$ versus $\vec p_{f3}^2$ from  15  $\bf_{6}^{12}C$ are shown in the top panel of Figure \ref{UC12}.   
The same values as a function of the $\Delta$  kinetic energy $T$ are shown on the bottom panel.   The extracted values of $U^{QE}_{opt}$ and
 $U_{opt}^\Delta$  versus $\vec p_{f3}^2$  (and $T$)  from  5  $\bf_{20}^{40}Ca$  spectra and one $\bf_{18}^{40}Ar$  spectrum are shown in the top and bottom panels of Fig. \ref{UCa}.  
Similarly, values extracted of $U^{QE}_{opt}$ and $U_{opt}^\Delta$ versus   $\vec p_{f3}^2$ (and $T$) for $\bf_{3}^{6}Li$, $\bf_{18}^{27}Al$, $\bf_{26}^{56}Fe$.
 are  shown in   Fig. \ref{remaining}.  For the  $\bf_{208}^{82}Pb$ spectra  the Fermi smearing is large and we only extract values of $U^{QE}_{opt}$ from the data. 

For the $\bf_{6}^{12}C$ nucleus, the values of  $U^\Delta_{opt}$  versus $\vec p_{f3}^2$ and $T^\Delta$ shown in Fig. \ref{UC12}  are fit to  linear functions which are shown as solid grey lines. The intercept  at $\vec p_{f3}^2=0$ and the slope of the fit to  $U^{\Delta}_{opt}$ versus $\vec p_{f3}^2$ as well as the  intercept and slope of the fit to  $U^{\Delta}_{opt}$ as a function of T  are also given in Table \ref{Table1}.  Fits for $\bf_{6}^{12}C$ using a different functional form are discussed in the Appendix and shown in Table \ref{Table2}.

As seen in Fig. \ref{UC12}, for the $\bf_{6}^{12}C$ data,  the linear fits to $U^{QE}_{opt}$  and  the linear fits to  $U^\Delta_{opt}$ cross zero at  approximately the same values of  $\vec p_{f3}^2$ (and $T$).  For  $\bf_{6}^{12}C$  we  have measurements of $U^\Delta_{opt}$  over a sufficient range of  $\vec p_{f3}^2$ and $T$ to perform a two parameter fit.   Because of the small number of measurements of $U^\Delta_{opt}$ for all the other nuclei, we do a one parameter fit for the slopes of  $U^\Delta_{opt}$ versus $\vec p_{f3}^2$ (and $T$) under the assumption that   $U^\Delta_{opt}$  crosses zero at the same values of  $\vec p_{f3}^2$ (and T) as the fits to $U^{QE}_{opt}$ for QE nucleons. The intercepts  at $\vec p_{f3}^2=0$ and the slopes of the fits to  $U_{opt}^\Delta$ versus $\vec p_{f3}^2$ (and $T$)  are also given in Table \ref{Table1}.   Note that the fits are only valid in the regions for which we have data.  We find that $U^\Delta \approx$1.5~$U^{nucleon}$ for $\bf_{6}^{12}C$. 
\subsection{Discussion of the optical potential for the the $\Delta$ resonance in the final state}
%
%
\subsubsection{Comparison to GiBUU}
%
As mentioned earlier,  GiBUU describes the initial state as a nucleon bound  in a potential $U$  which depends on both the local density $\rho$  and momentum.  The same density and momentum dependent  potential is used for the initial and final state nucleon. 

  For the case of the production of the $\Delta$ resonance, the GiBUU formalism requires a density and momentum dependent potential for the $\Delta$ resonance in the nucleus.  What is used\cite{GiBUU2} in GiBUU is  $U^\Delta_{GiBUU}$=~(2/3)~$U^{nucleon}_{GiBUU}$. 
 This is not in agreement with our results which indicate that $U^\Delta \approx$ 1.5~$U^{nucleon}$ for $\bf_{6}^{12}C$. 
 
 The short lifetime of the $\Delta$ ($5.63 \times 10^{-24}$ sec) implies that for the low energy transfers discussed in this paper, the $\Delta$ decays occur inside the nucleus. Consequently, one would expect that the optical potential for the $\Delta$ should reflect the sum of the corresponding optical potentials of the decay nucleon and pion. This is consistent with our results which show that $U^\Delta$ more negative than $U^{QE}$.
.
   
\subsubsection{ Effective mass of nucleons and $\Delta$ resonances in the nuclear medium}
  
Some authors\cite{miller} have cast the effect of the nuclear optical potential on the nucleon and $\Delta$(1232)  as an energy dependent change in their effective mass in the nuclear medium.  Under this interpretation, both the nucleon and the $\Delta$ revert back to their free mass values  after leaving the nucleus.  For example, the distribution of the  final state mass of the decay particles of  $\Delta$ resonances  produced in neutrino-(Propane/Freon) interactions\cite{skat} peaks around 1.232 GeV. 

At low kinetic energy T,  both optical potentials are negative and therefore  can be interpreted as a $reduction$ in the effective masses of nucleons and $\Delta$ resonances when  produced in a nuclear medium \cite{miller}. Additional details are discussed in the Appendix.

However, as discussed in the next section the effective mass representation is not the approrpiate representation for MC generators such as \genie~and  \neut.

  \subsubsection{Structure functions of bound nucleons}
  %
In most impulse approximation Monte Carlo generators, the structure functions of the nucleus are expressed in terms of a convolution of the nucleon momentum distributions with the structure functions of bound nucleons.  The structure functions of bound nucleons are identified with the structure functions of free nucleons expressed in terms of $Q^2$ and the final state invariant mass $W$.  When the final state invariant mass is the the mass of the nucleon, the free nucleon form factors are used.  When the mass of the final state is $W$, the inelastic free nucleon structure functions for the corresponding $W$ and $Q^2$ are used\cite{BodekW, BodekRitchie}.  Consequently, the interaction of the final state of mass $W$ with the mean field of the nucleus expressed in terms of an average optical potential is more consistent with how  the structure functions of bound nucleons are related to structure functions of free nucleons  in these Monte Carlo generators,

\section {Extraction of  neutrino oscillations  parameters}
  \subsection{ Interaction energy}
  %
In the off-shell formalism of Bodek and Ritchie\cite{BodekRitchie}, which is used in \genie,~equations \ref{QE_equation} and \ref{Delta_Eq} can be written in terms of an energy dependent {\it interaction energy}: 
 \begin{fleqn}
 \begin{eqnarray}
  \label{eq_summary}
&&\nu +(M_{P,N} -\epsilon_{QE-interaction}^{off-shell P,N})= \sqrt{\vec p_{f3}^2+M_{P,N}^2}\\
&&\nu +(M_{P,N} -\epsilon_{\Delta-interaction}^{off-shell P,N})= \sqrt{\vec p_{f3}^2+W_{\Delta+,0}^2}~,\nonumber
\end{eqnarray}
 \end{fleqn}
where
\begin{fleqn}
 \begin{eqnarray} 
&&\epsilon_{QE-interaction}^{off-shell~ P,N}=\epsilon^{P,N}+ U^{QE}_{opt}(\vec p_{f3}^2)+|V_{eff}^{P,N}|\\
&&\epsilon_{\Delta-interaction}^{off-shell ~P,N}=\epsilon^{P,N}+ U^{\Delta}_{opt}(\vec p_{f3}^2)+|V_{eff}^{\Delta +,0}|.
\nonumber 
\end{eqnarray} 
  \end{fleqn}
  For electron scattering on a nucleon bound in  $\bf_{6}^{12}C$, our results imply that the {\it interaction energies} for the range 
of final state baryon kinetic energies between  0.05 and 0.3 GeV vary from 5 to 28 MeV for the nucleon and from 11 to 29 MeV for the $\Delta$.

In the  on-shell formalism of Moniz et al.\cite{Moniz} (used in \textsc{neut}) the {\it Moniz interaction energies} are defined as:
 \begin{fleqn}
 \begin{eqnarray}
&&\nu +(M_{P,N}+T_i^{P,N} -\epsilon_{QE-interaction}^{Moniz~P,N})= \sqrt{\vec p_{f3}^2+M_{P,N}^2}\nonumber\\
&&\nu +(M_{P,N})+T_i^{P,N} -\epsilon_{\Delta-interaction}^{Moniz~P,N}= \sqrt{\vec p_{f3}^2+W_{\Delta+,0}^2}~,\nonumber
\end{eqnarray}
 \end{fleqn}
where  $T_i^{P,N}$ is the kinetic energy of the initial state nucleon (which  on average is equal to $\frac{3}{5}\frac {K_F^{P,N})^2}{2M^{P,N}}$ for a Fermi gas with Fermi momentum $K_F^{P,N}$).

 Comparing to  Equation  \ref{QE_equation} we obtain
\begin{fleqn}
 \begin{eqnarray} 
&& \epsilon_{QE-interaction}^{Moniz~P,N}=\epsilon^{P,N}+T_i^{P,N}+ U^{QE}_{opt}(\vec p_{f3}^2)+|V_{eff}^{P,N}|\nonumber \\
&& \epsilon_{\Delta-interaction}^{Moniz~P,N}=\epsilon^{P,N}+T_i^{P,N}+ U^{\Delta}_{opt}(\vec p_{f3}^2)+|V_{eff}^{\Delta +,0}|.
 \nonumber 
\end{eqnarray} 
  \end{fleqn}
 For electron  scattering on a  nucleon bound in $\bf_{6}^{12}C$, our results imply that the {\it Moniz interaction energies} for the range 
of final state baryon kinetic energies between  0.05 and 0.3 GeV vary from 21 to 44 MeV for the nucleon and from 4 to 44 MeV for the $\Delta$.

 However, in the analysis of  Moniz et al.\cite{Moniz} the {\it interaction energies}
 $\epsilon_{QE-interaction}^{Moniz~P,N}$ and $\epsilon_{\Delta-interaction}^{Moniz~P,N}$  are  assumed to be the same which we find is not correct. In addition, the two interaction energies as defined by Moniz are  assumed to be constant,  which we also find is not correct (these interaction energies depend on both  the initial state kinetic energy and on $U_{opt}$ which is a function of  kinetic energy of the final state baryon).
 %
 
  \subsection {Reducing systematic error in the measurements of neutrino oscillations 
  parameters}
As shown above, the {\it interactions energies} for the nucleon and the $\Delta$ are different and over the range of final state baryon kinetic energies from 0.05 to 0.3 GeV they change by about 20 MeV.  Using our determinations of the removal energies and energy dependent optical potentials  reduces the systematic error in the {\it interaction energies} for QE-like events from  $\pm$ 20 MeV  to $\pm$ 5 MeV.  

In reference \cite{optpaper} we estimate  that a +20 MeV change in the interaction energy used in the MC corresponds to a change in  $\Delta{m_{32}^2}$  of $+0.03 \times 10^{-3}~\rm eV^2$, which is the $largest$ contribution to the total systematic error in  $\Delta{m_{32}^2}$.  This estimate is consistent with the estimate of the  \textsc{t2k} collaboration which 
 reports\cite{t2k-impact} that    ``for the statistics of the 2018 data set, a shift of 20 MeV in the binding energy parameter introduces a bias of 20\% for  $\sin^2\theta_{23}$  and 40\% for $\Delta{m_{32}^2}$  with respect to the size of the systematics errors, assuming maximal  $\sin^2\theta_{23}$''.
Consequently in neutrino oscillations experiments our measurements can reduce the systematic uncertainty  in the reconstruction of the neutrino energy primarily in experiments such as T2K and Hyper-K\cite{K2K}  that infer the energy of the neutrino from the kinematics of the final state lepton. 
  \section{Conclusions}
%
We report on the   extraction (from electron scattering data)  of the average nuclear optical potentials for both  nucleons  and $\Delta$(1232) resonances in the final state as a function of kinetic energy of the final state baryons.  The data  show that:

\begin{enumerate}
\item {\it Nucleons optical potential}:   The measurements of the average optical potential for a final state nucleon  $U^{QE}_{opt}$  for $\bf_{3}^{6}Li$ and $\bf_{26}^{56}Fe$ are in good  agreement with calculations based on the  Cooper 1993\cite{Cooper1993}  and Cooper 2009\cite{Cooper2009} formalisms. The measurements are more negative than the  calculations for $\bf_{6}^{12}C$+$\bf_{8}^{16}O$, $\bf_{18}^{27}Al$, and  $\bf_{20}^{40}Ca$+$\bf_{18}^{40}Ar$, and the measurements are less negative than the calculations for $\bf_{82}^{208}Pb$. For the $\bf_{6}^{12}C$ nucleus,  although both   Cooper calculations  of  $U^{QE}_{opt}$ are above the data, the Cooper 1993\cite{Cooper1993} calculations are closer to the data than the Cooper 2009\cite{Cooper2009} calculations.   We provide fits to the nucleon optical potentials for use in modeling of QE neutrino scattering on nuclear targets. 
\item {\it $\Delta$ optical potential}:  We find that the average optical potential for a $\Delta$ resonance in the final state $U_{opt}^\Delta$   is more negative than the average optical  potential for a final state nucleon  with  $U_{opt}^\Delta \approx$1.5~$U^{QE}_{opt}$ for $\bf_{6}^{12}C$. This is different from the optical potential used in GiBUU\cite{GiBUU2} for which $U^{\Delta}_{GiBUU}$=(2/3)$U^{nucleon}_{GiBUU}$ is assumed.  We provide fits to the nucleon optical potentials for use in modeling  $\Delta$ resonance production in neutrino scattering on nuclear targets. 
\item {\it Modeling QE-like events}: Using the measurements of these four  parameters  $\epsilon^{P,N}$, $U^{QE}_{opt}$, $U^\Delta_{opt}$,  and  $V_{eff}$  we can model the energies of  leptons, nucleons   and $\Delta$ resonances in the final state for QE-like events.  For neutrino oscillations experiments these measurements can reduce the systematic uncertainty  in the reconstruction of the neutrino energy primarily in experiments such as T2K and Hyper-K\cite{K2K}  that infer the energy of the neutrino from the kinematics of the final state lepton.  
\end{enumerate}
\begin{figure} 
\centering
      \includegraphics[width=3.3in,height=2.1in]{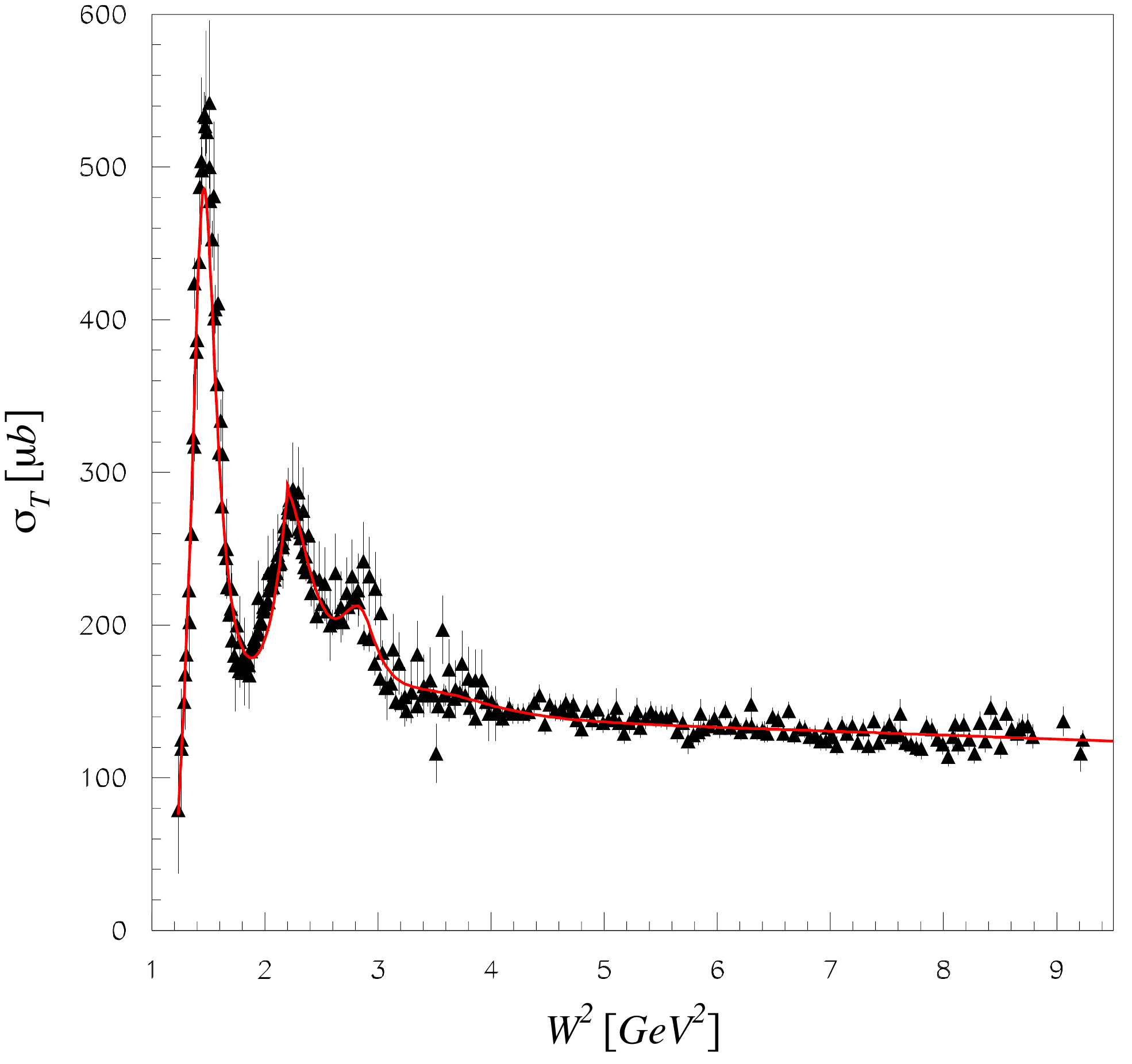}
    \includegraphics[width=3.3in,height=2.6in]{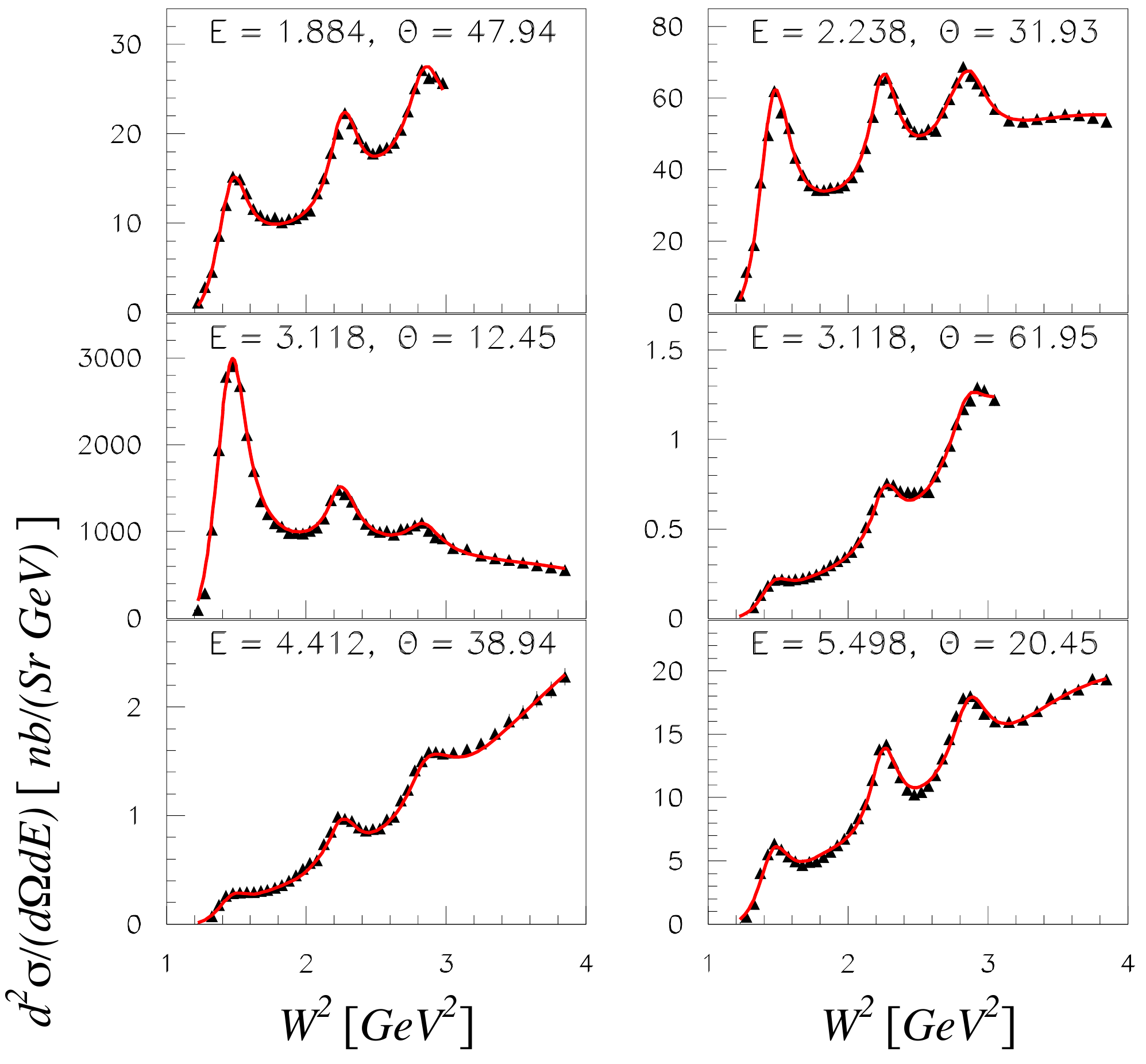}
  \caption{
\footnotesize\addtolength{\baselineskip}{-1\baselineskip} 
Examples of Bosted-Christy fits to photoproduction data on protons and Jefferson lab electron-proton cross sections in the resonance region. Curves are from reference\cite{eric}.}
\label{Bosted_Christy}
\end{figure} 
\section{Appendix}
\subsection{Bosted-Christy fits to nucleon inelastic structure functions}
\label{Bosted_Christy}
%
 In the calculation of  the inelastic  cross section for the production of resonances including the continuum we use the Bosted-Christy fits\cite{eric} to the inelastic structure functions for protons and neutrons. These  are fits  to a wide range of inelastic electron scattering data on protons and deuterons including photo-production data at $Q^2$=0. 
The fits  describe the inelastic structure functions for protons and neutrons including resonances and continuum over a wide range of $Q^2$. A comparison of  Bosted-Christy fits\cite{eric} to  photo-production data on protons and to a few examples of electron-proton cross sections measured at Jefferson Lab are shown in Figure \ref{Bosted_Christy}.

The width of the $\Delta$ produced in photo-production\cite{DAPHNE} on free nucleons  as well as at  very low $Q^2$ electron scattering on free nucleons is smaller than at larger values of $Q^2$.  This leads to an $apparent$ reduction of the location of the $\Delta$ peak in $W$ from 1.232 GeV to $\approx$1.220 GeV at very low $Q^2$.  This change in the width (but keeping the mass of the $\Delta$ at 1.232 GeV) is taken into account in  Bosted-Christy fits\cite{eric} that we use to parametrize the electro production of resonances on free nucleons.  Our analysis  also includes  the effects of Fermi motion on both the peak location and width of the $\Delta$ when produced in the nucleus.
\begin{figure*} 
\centering
      \includegraphics[width=3.5in,height=2.5in]{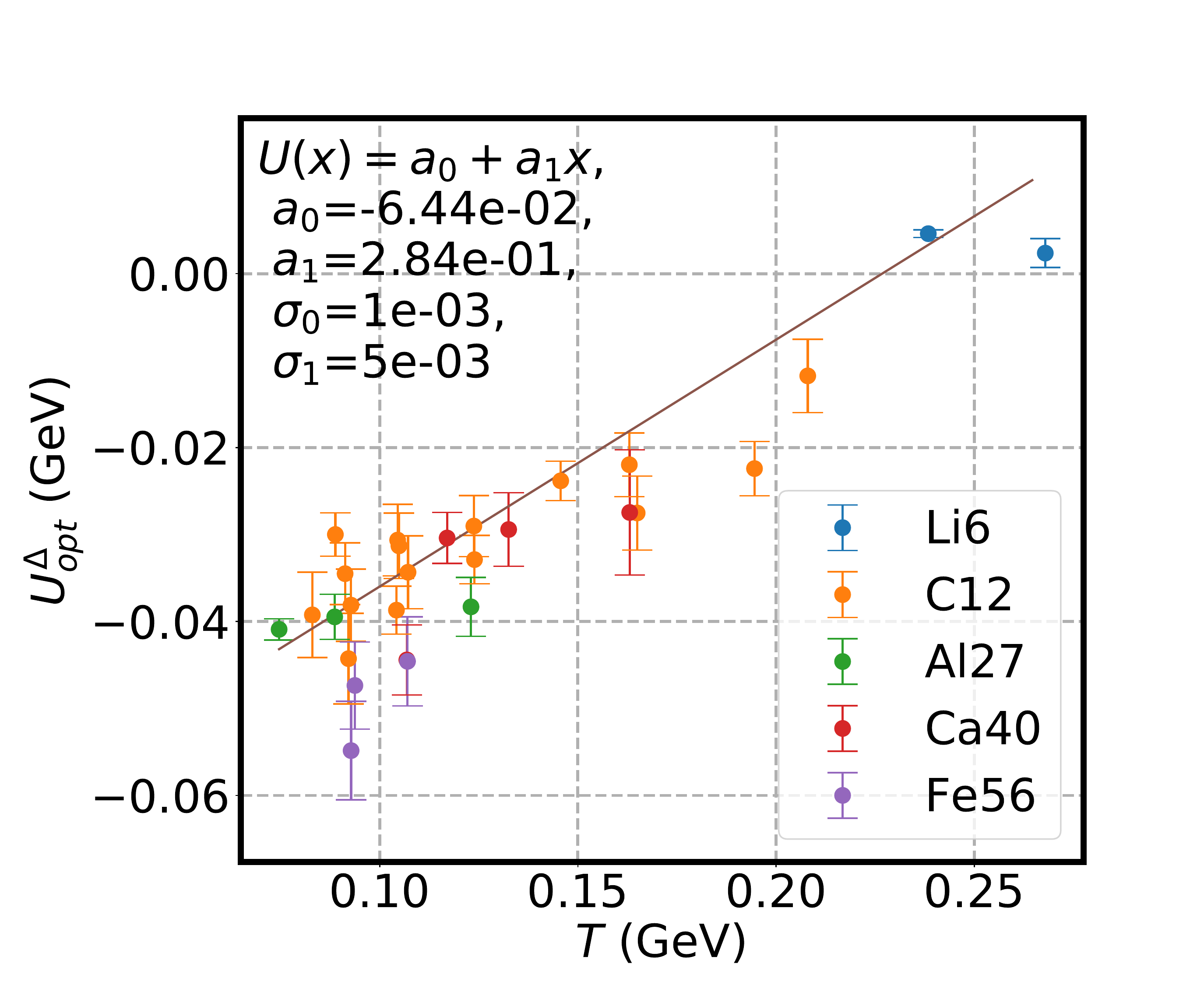}
    \includegraphics[width=3.4in,height=2.5in]{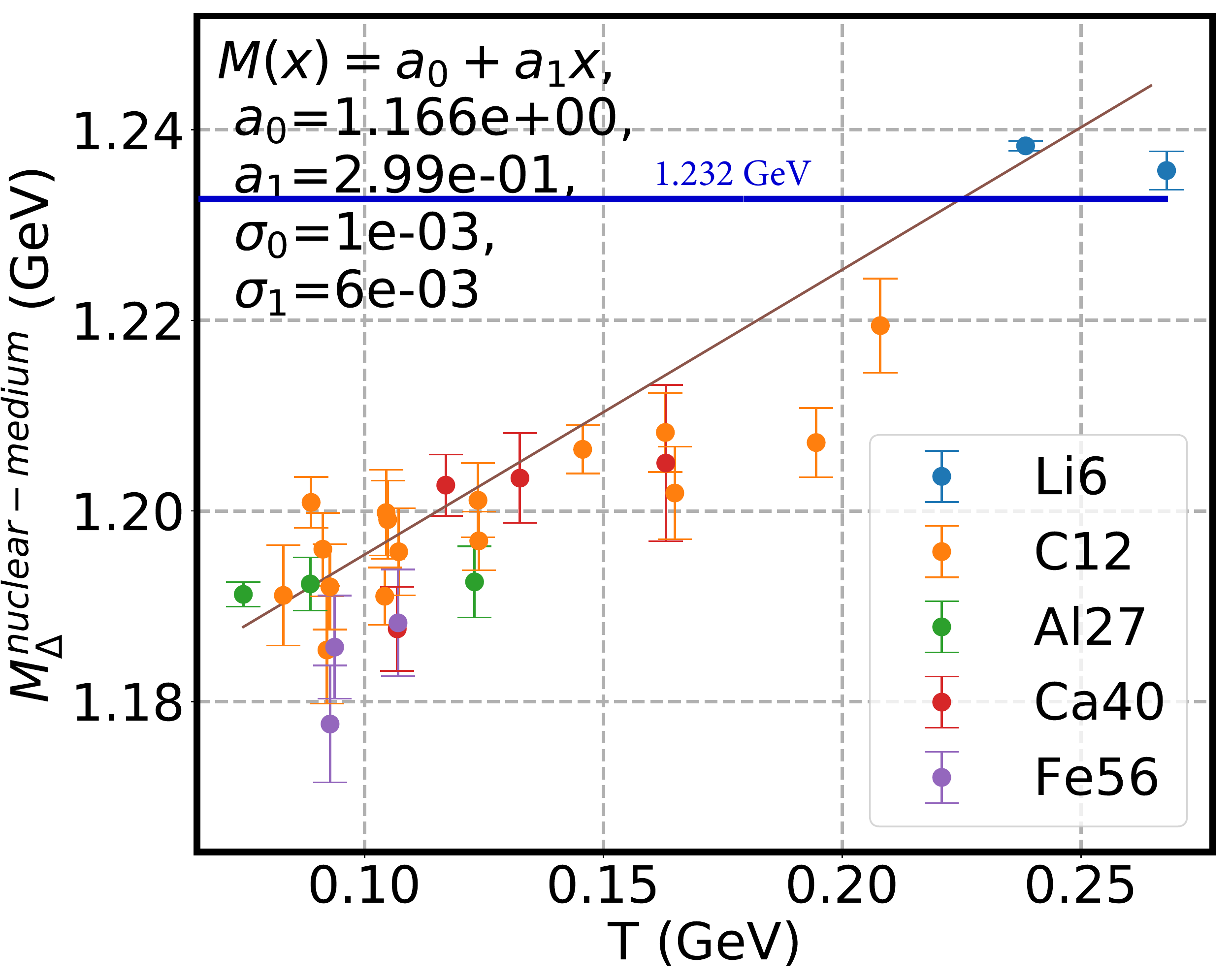}
       \includegraphics[width=3.7in,height=2.5in]{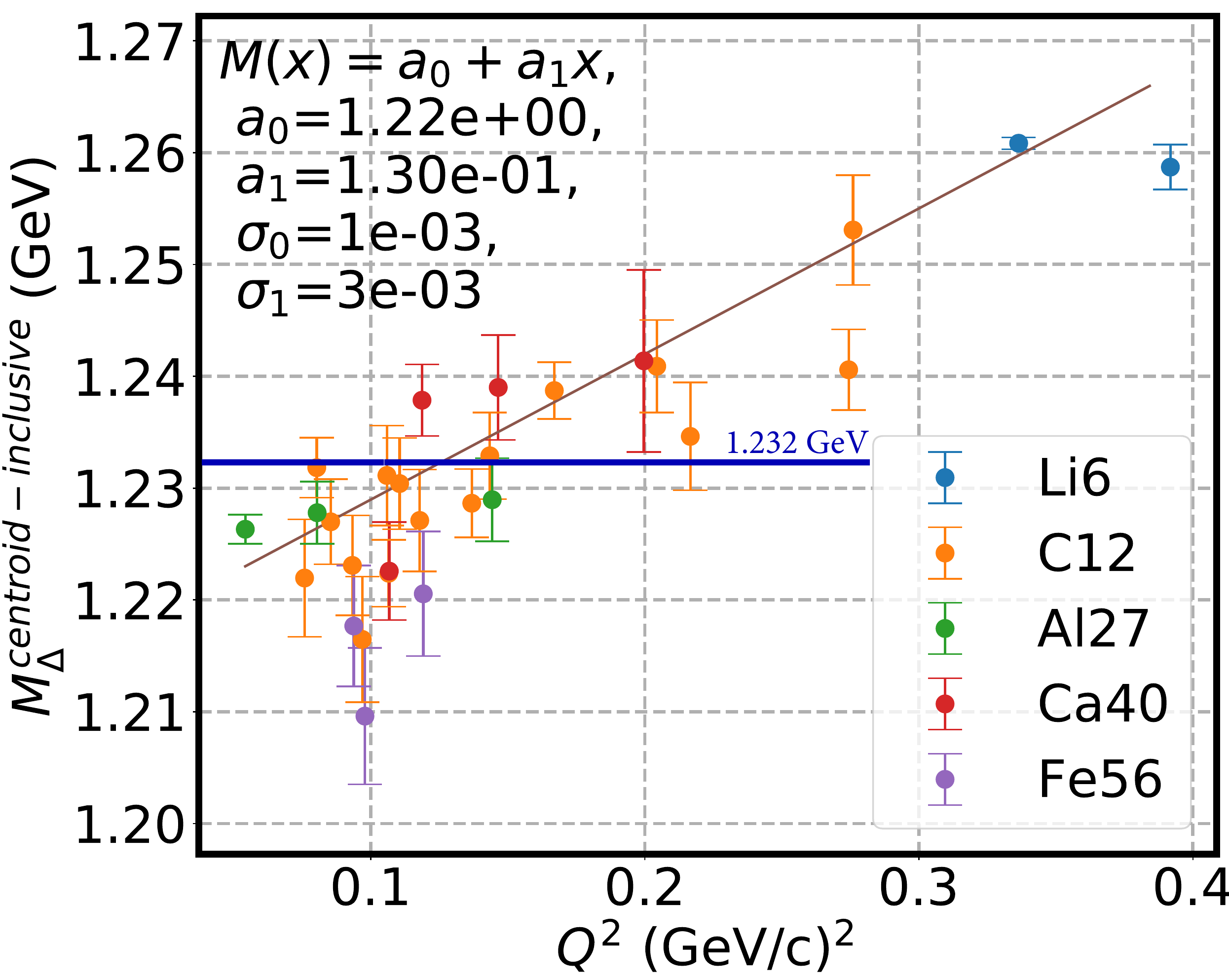}
                 \includegraphics[width=3.2in, height=2.5in]{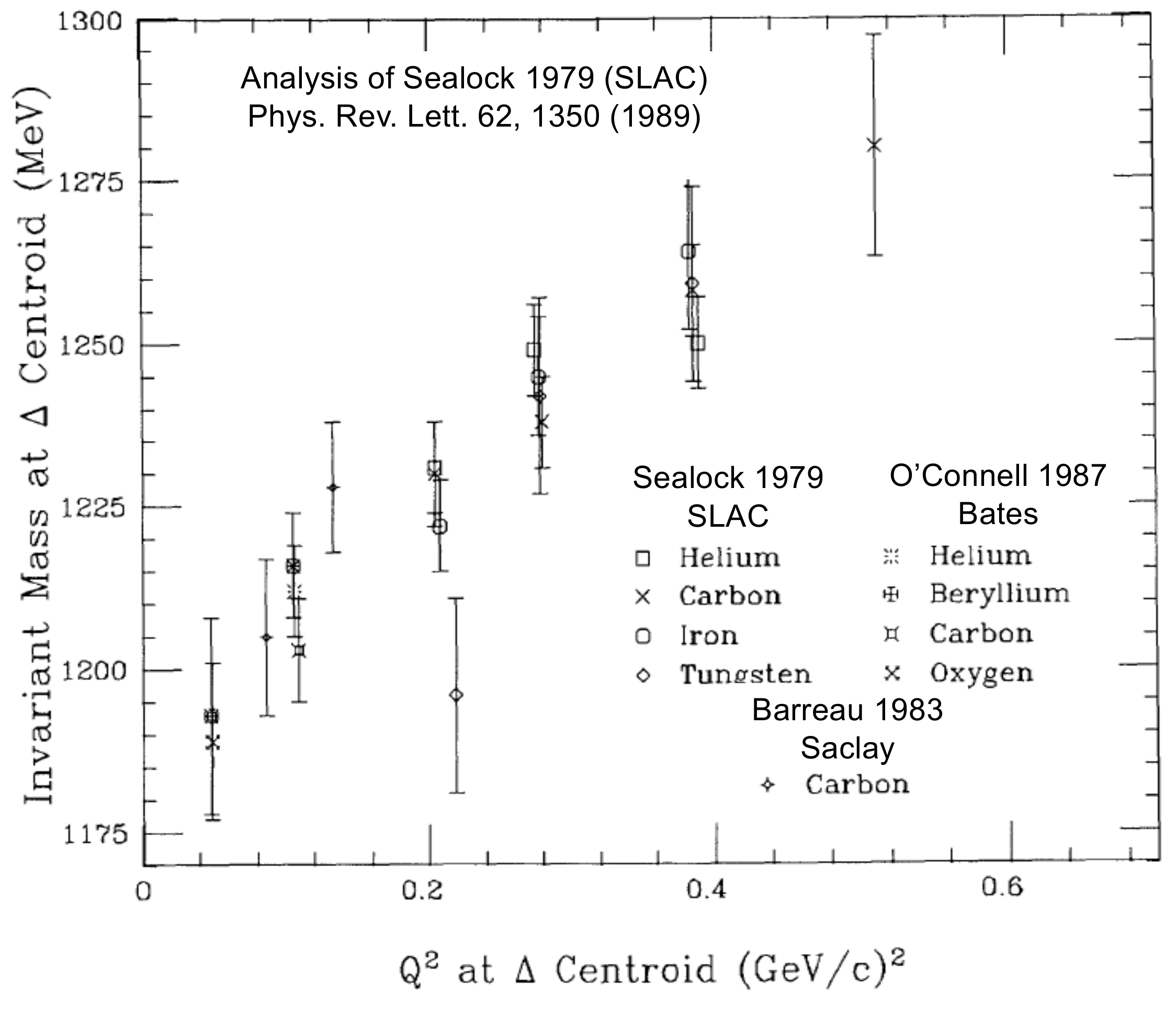}
  \caption{
\footnotesize\addtolength{\baselineskip}{-1\baselineskip} 
  (a) The top left panel shows values of $U_{opt}^{\Delta}$ versus kinetic energy of the $\Delta$ resonance in the final state (for all nuclei). (b)  The top right panel shows values of $M_{\Delta}^{nuclear-medium}$, the effective average $\Delta$ mass while inside the nucleus, versus the  kinetic energy of the $\Delta$ (for all nuclei). (c)  The bottom left panel shows values of   $M_{\Delta}^{centroid-inclusive}$ (the centroid of the peak in invariant mass W for the inclusive electron scattering in the $\Delta$ mass region) shown versus $Q^2$ (for all nuclei). (d)  The bottom right panel shows values of  $M_{\Delta}^{centroid-inclusive}$ versus $Q^2$ for various nuclei from the paper by Sealock et.al\cite{Sealock:1989nx}.  Here,  the Bates points are from  O'Connell et al.  \cite{O'Connell:1987ag}, and the Saclay points are from  Barreau et al. \cite{Barreau:1983ht}.  Our results indicate that the change in $M_{\Delta}^{centroid-inclusive}$ versus $Q^2$ originates primarily from the dependence of the average optical potential $U_{opt}^{\Delta}$ on the kinetic energy of the $\Delta$ resonance in the final state.}
  \label{all_nuclei}
\end{figure*}
\subsection{The effective mass of the nucleon and $\Delta$(1232) in the nuclear medium}
\label{average_mass}
%
%
For  purpose of comparison to other publications, we transform our results for the average optical potential for the $\Delta$  to an equivalent change of the effective mass of the $\Delta$ in the nucleus using the following expression with $M_{\Delta}^{free}$ = 1.232 GeV.
\begin{fleqn}
 \begin{eqnarray}
  && \sqrt{ (\vec k +\vec q_3)^2+(M_{\Delta}^{nuclear-medium})^2 } = \nonumber \\
  && \sqrt{ (\vec k +\vec q_3)^2+(M_{\Delta}^{free})^2} +U^{\Delta}_{opt}+  |V_{eff}|.
 \end{eqnarray} 
  \end{fleqn}
  
 Similarly,  the peak in the inclusive distribution in $W$ $M_{\Delta}^{centroid-inclusive}$(but not including the effect of Fermi motion) can be extracted from the following expression
 \begin{fleqn}
 \begin{eqnarray}
  && \sqrt{ (\vec k +\vec q_3)^2+(M_{\Delta}^{centroid-inclusive})^2} = \nonumber \\
  && \sqrt{ (\vec k +\vec q_3)^2+(M_{\Delta}^{free})^2} +U^{\Delta}_{opt}+  |V_{eff}|+\epsilon.
 \end{eqnarray} 
  \end{fleqn}
In the above expressions we use the average $\epsilon$ and  $|V_{eff}|$ for neutrons and protons. In order to compare to the analysis of Sealock et.al\cite{Sealock:1989nx} we show the data for all nuclei on a single plot in Figure \ref{all_nuclei}.

The top left panel of Fig. \ref{all_nuclei} shows values of $U_{opt}^{\Delta}$ versus kinetic energy of the $\Delta$ resonance in the final state  (for all nuclei).  The points can be approximated by the following expression 

\vspace {0.05 in}
\noindent $U_{opt}^{\Delta}$(GeV) $\approx$

  (-0.0644$\pm$0.0010) +  (0.284$\pm$0.005) $T^\Delta$(GeV). 
\vspace {0.05 in}

 The top right panel shows  $U_{opt}^{\Delta}$ as  $M_{\Delta}^{nuclear-medium}$, the  average  $\Delta$ effective mass inside the nucleus, versus the  kinetic energy of the $\Delta$ (for all nuclei).  The points can be approximated by the following expression   
 
\vspace {0.05 in}
 \noindent $M_{\Delta}^{nuclear-medium}$(GeV)  $\approx$
 
   (1.166$\pm$0.001) +  (0.299$\pm$0.006) $T^\Delta$(GeV).  
 \vspace {0.05 in}
 
 The bottom left panel shows the centroids of the peak in invariant mass W for the inclusive electron scattering in the $\Delta$ mass region ($M_{\Delta}^{centroid-inclusive}$) versus $Q^2$ (for all nuclei).  The points can be approximated by the expression 
 
 \vspace {0.05 in}
 \noindent  $M_{\Delta}^{centroid-inclusive}$ $\approx$
 
  (1.220$\pm$0.001) +  (0.130$\pm$0.03) $Q^2$(GeV$^2$). 
  \vspace {0.05 in} 
 
 The bottom right panel shows $M_{\Delta}^{centroid-inclusive}$ versus $Q^2$ for various nuclei from the paper by Sealock et.al\cite{Sealock:1989nx}.  Here,  the Bates points are from  O'Connell et al.  \cite{O'Connell:1987ag}, and the Saclay points are from  Barreau et al. \cite{Barreau:1983ht}.  
 
 Note that unlike our analysis, the $M_{\Delta}^{centroid-inclusive}$ values extracted  by Sealock et.al.  do not correct for the apparent shift in the centroid from the known decrease of the width of the $\Delta$ at low $Q^2$ (which shift the apparent centroid to lower mass), nor do they correct for the effect of Fermi motion. 
 In addition, the Sealock et.al\cite{Sealock:1989nx} analysis includes spectra which have a much  larger contribution from the continuum than the spectra used in our analysis.
 
    We conclude that  the change in $M_{\Delta}^{centroid-inclusive}$ versus $Q^2$  originates primarily  from the dependence of the average optical potential $U_{opt}^{\Delta}$ on the kinetic energy of the $\Delta$ resonance in the final state.
    \subsection{Comparison to the analysis of O'Connell and Sealock}
    \label{OConnel_analysis}
  %
   A previous  extraction  of the average nucleon and $\Delta$ potentials from electron scattering cross sections on  $\bf_{6}^{12}C$   was published by O'Connell and Sealock\cite{early} in 1990. 
 They  find that the potential for the $\Delta$ is more negative than the potential for the nucleon with $U^\Delta \approx$ 2.5~$U^{nucleon}$ for $\bf_{6}^{12}C$.  Although qualitatively their conclusions are similar to ours,  there are significant differences between the the two analyses
   
\begin{figure} 
\centering
    \includegraphics[width=3.5in,height=2.6in]{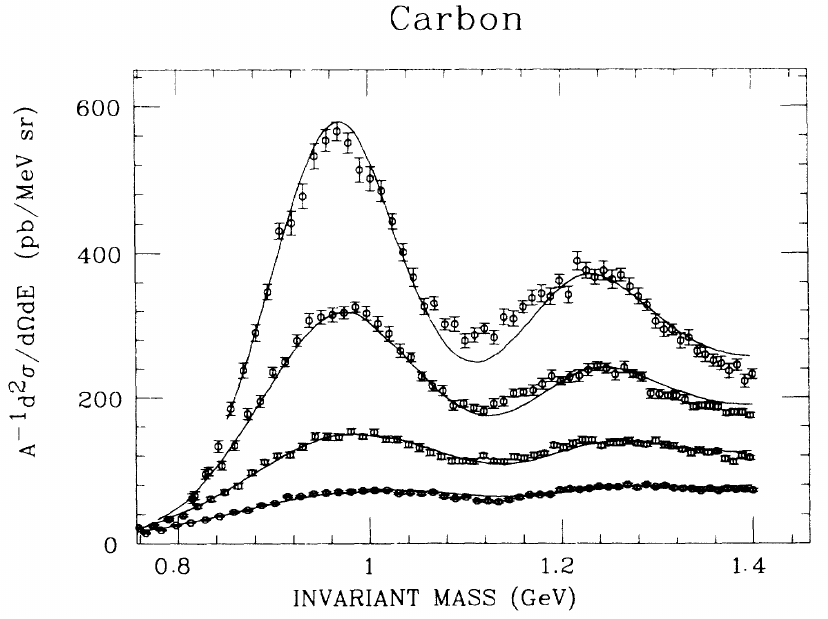}
    \includegraphics[width=3.5in,height=3.0in]{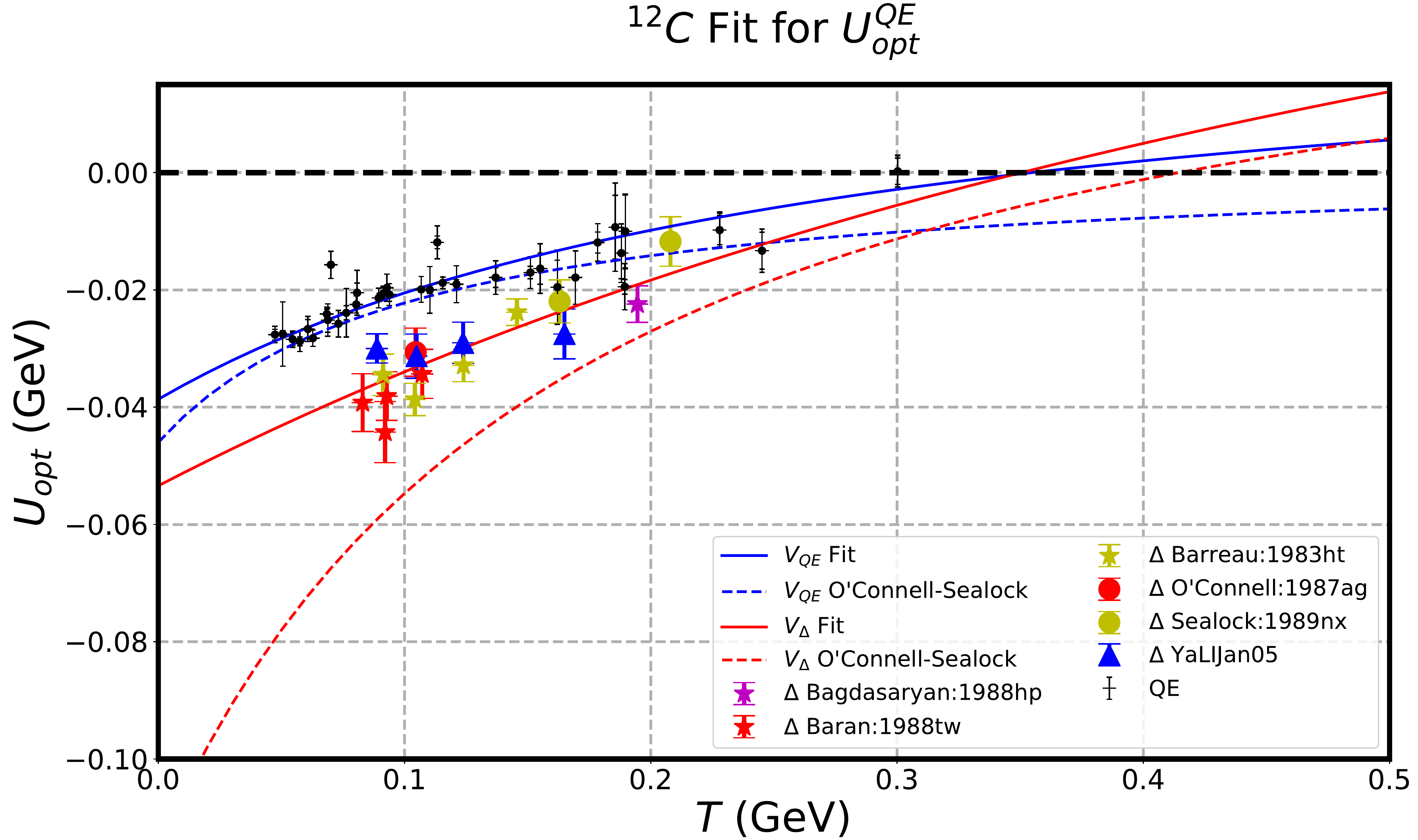}
  \caption{
\footnotesize\addtolength{\baselineskip}{-1\baselineskip} 
Top panel: The differential cross sections (as a function of final state invariant mass) used in the O'Connell and Sealock\cite{early}  analysis on $\bf_{6}^{12}C$. All these data were taken at a spectrometer angle of 37.5$^0$. The beam energies from top to bottom are 0.96, 1.1, 1.3 and 1.5 GeV.   We note that the contributions from the continuum is much larger than for the cross sections used in our analysis.
Bottom panel: Comparison of  the O'Connell-Sealock fits for $U^{QE}_{opt}$ and $U^{\Delta}_{opt}$ for $\bf_{6}^{12}C$  to our data.  Our fits
  (using the same functional form $U ={V_0}/{(1+p^2/p_0^2)} + V_1$) are shown as solid lines and the  O'Connell-Sealock fits are shown as dashed lines.
 }
\label{OConnell_mass}
\end{figure}

As shown in the top panel of Fig. \ref{OConnell_mass} the spectra used in the analysis of O'Connell and Sealock have much  larger contributions from the continuum than the spectra used in our analysis.   Consequently,  we believe that their results for the average potential for the $\Delta$ resonance  should have an additional model uncertainty.  Nonetheless, comparisons between their results and our results for the optical potentials for the nucleon and the $\Delta$ are discussed below,

In the   O'Connell and Sealock analysis the nucleon optical potential is used for both the initial state and final state nucleons.  A specific functional form is assumed for both the nucleon and $\Delta$ nuclear potentials.  Equations   \ref{eq2} and \ref{QE_equation} are not used in the O'Connel-Sealock analysis.  Instead the following expression for QE scattering is used
   \begin{fleqn}
 \begin{eqnarray}
 &&\nu +  \sqrt{ (\vec p_i^2 +M_{P,N}^2} +U_{O'Connell}^{nucleon}(p_i^2) = \nonumber \\
  && ~~~~~~~~~~~~~~~~\sqrt{ (\vec k +\vec q_3)^2+M_{P,N}^2} +U_{O'Connell}^{nucleon}(p_f^2),\nonumber
  \end{eqnarray} 
  \end{fleqn}
 and the following expression is used for inelastic scattering with a final state invariant mass W. 
\begin{fleqn}
 \begin{eqnarray}
  &&\nu +  \sqrt{ (\vec p_i^2 +M_{P,N}^2} +U_{O'Connell}^{nucleon}(p_i^2) = \nonumber \\
  &&~~~~~~~~~~~~~~~~ \sqrt{ (\vec k +\vec q_3)^2+W^2} +U_{O'Connell}^{\Delta}(p_f^2),\nonumber
 \end{eqnarray} 
  \end{fleqn}  
   where  $p_i^2=K^2_{F}/2$. Both nuclear potentials are then fit to the following functional form
    \begin{eqnarray}
    \label{new_fit}
  U &=&\frac{V_0}{1+p^2/p_0^2} + V_1. 
  \end{eqnarray}
    However, since  removal energy information is not used in the  O'Connell and Sealock analysis,  $U^{nucleon}(p_i^2=K^2_{F}/2)$ is not  constrained by the differential cross sections. Consequently, for the nucleon they set $V_1^{nucleon}$  such that the fit  to the nucleon potential yields $U^{nucleon}(p_i^2=K^2_{F}/2)$= 41 MeV.   
    
Comparing their expressions to equation \ref{QE_equation} we can estimate 
     $$U^{P,N}(p_i^2)=\epsilon^{P,N}+T_i^{P,N}$$
     which yields 40.4 MeV for  the proton and 43.0 MeV for the neutron for $p_i^2=(1/2)K^2_{F}$ (T=12.5 MeV), and 43.0 MeV for  the proton and 45.6 MeV for the neutron for $p_i^2=(3/5)K^2_{F}$ (T= 15 MeV), which is the average kinetic energy for a Fermi gas.  Therefore, their assumption that $U^{nucleon}(p_i^2=K^2_{F}/2)$= 41 MeV is consistent with our values of the removal energies within an uncertainty of 3 MeV.

 %
 \begin{table}
 \begin{center}					
\begin{tabular}{|cc|ccc|}					
\hline 					
Analysis	&  Final & $V_0$  & $V_1$  & $p_0$\\
 &state & MeV  & MeV  & GeV \\
\hline
This  & N & -62$\pm$2  & 23$\pm$4 &0.691$\pm$0.049 \\
analysis  & $\Delta$ & -138$\pm$6 & 85 &1.253$\pm$0.117 \\
 \hline
OConnell+  & N & -46$\pm$6  & 0 &0.430$\pm$0.100 \\
Sealock  & $\Delta$ & -153$\pm$22 & 38$\pm$3  & 0.628$\pm$0.088 \\
  \hline  
\end{tabular}
\caption{  Comparison of our fits to $U^{QE}_{opt}$ and $U^{\Delta}_{opt}$ for $\bf_{6}^{12}C$  to  the O'Connell-Sealock fits using the same functional form $U ={V_0}/{(1+p^2/p_0^2)} + V_1$.  
}	
\label{Table2} 					
\end{center}					
\end{table}

  For  comparison, we also fit  our values for the average optical potentials for the nucleon and $\Delta$ to the functional form given in equation \ref{new_fit}.
  For the nucleon, there is sufficient data to extract all of the fit parameters. For the $\Delta$ we set $V_1=85$ MeV, for which the potentials for the $\Delta$ and the nucleon cross zero at the same value of kinetic energy T.  
     Comparison of  the O'Connell-Sealock fits for $U^{QE}_{opt}$ and $U^{\Delta}_{opt}$ for $\bf_{6}^{12}C$  to our data  are shown in the bottom panel of Fig.  \ref{OConnell_mass}.   Our fits
  (using the same functional form $U ={V_0}/{(1+p^2/p_0^2)} + V_1$) are shown as solid lines and the  O'Connell-Sealock fits are shown as dashed lines. 
     
     A comparison of the fit parameters extracted in our analysis, and the parameters from O'Connell and Sealock are given in Table \ref{Table2}.  Note that the O'Connell-Sealock fits are only valid in the region  of $p_f^2$ between 0.16 and 1.0 GeV$^2$ ($0.08 < T^N\ < 0.43$, $0.06 < T^\Delta < 0.36$).  As can be seen in the bottom panel of  Fig. \ref{OConnell_mass}  the O'Connell-Sealock fit for the average potential for the nucleon is in good agreement with our data.
      
    However, the O'Connell and Sealock fit for the average potential for the $\Delta$ is much more negative than our data.  And, as mentioned before,  the spectra used in the analysis of O'Connell and Sealock (top panel of Fig. \ref{OConnell_mass}) have much  larger contributions from the continuum than the spectra used in our analysis.   Consequently, we believe that their extractions of the average potential for the $\Delta$ resonance have additional model uncertainties.


%
\end{document}